\documentclass[
reprint,
superscriptaddress,
amsmath,amssymb,
aps,
prl,
floatfix
]{revtex4-2}

\usepackage{graphicx}% Include figure files
\usepackage{dcolumn}% Align table columns on decimal point
\usepackage{bm}% bold math
\usepackage[colorlinks,citecolor=magenta]{hyperref}

\begin{document}

\title{Emitter-Optomechanical Interaction in Ultra-High-Q hBN Nanocavities}

\author{Chenjiang Qian}
\email{chenjiang.qian@wsi.tum.de}
\author{Viviana Villafañe}
\author{Martin Schalk}
\affiliation{Walter Schottky Institut and Physik Department, Technische Universit{\" a}t M{\" u}nchen, Am Coulombwall 4, 85748 Garching, Germany}
\author{Georgy V. Astakhov}
\author{Ulrich Kentsch}
\author{Manfred Helm}
\affiliation{Helmholtz-Zentrum Dresden-Rossendorf, Institute of Ion Beam Physics and Materials Research, 01328 Dresden, Germany}
\author{Pedro Soubelet}
\author{Andreas V. Stier}
\affiliation{Walter Schottky Institut and Physik Department, Technische Universit{\" a}t M{\" u}nchen, Am Coulombwall 4, 85748 Garching, Germany}
\author{Jonathan J. Finley}
\email{finley@wsi.tum.de}
\affiliation{Walter Schottky Institut and Physik Department, Technische Universit{\" a}t M{\" u}nchen, Am Coulombwall 4, 85748 Garching, Germany}
%\date{\today}

\begin{abstract}
    Integrating quantum emitters into nanocavities which simultaneously couple to the photonic and mechanical modes is critical for interfacing electron spins, photons and phonons in the cavity QED system.
    Here, we investigate the interaction between the charged boron vacancy $V_B^-$, ultra-high-Q ($\sim10^5$) cavity photonic modes and local phonon modes.
    A pronounced asymmetry is observed in the emission spectrum for cavities with Q-factor above a threshold of 10$^4$.
    Similar asymmetries are not observed for cavities without $V_B^-$ centers.
    To explain our findings, we model the system with phonon-induced light-matter coupling based on $V_B^-$ centers, and compare to the Jaynes-Cummings model for usual emitters.
    Our results reveal that the multipartite interplay arises during the light-matter coupling of $V_B^-$ centers, illustrating that it is phonon-induced, rather than being caused by thermal population of phonon modes. 
    Such emitter-optomechanical interaction between different photon ($V_B^-$ emission, cavity photonic) and phonon ($V_B^-$ phonon, cavity mechanical) modes provides a novel system to interface spin defects, photons and phonons in condensed matters.
\end{abstract}

\maketitle

% \section{\label{sec1}Introduction}

Light-matter coupling between one or more quantum emitters and the photonic mode in a nanocavity is the central pillar of cavity QED experiments.
This coupling provides the spin-photon interface as the basic building block from which nanophotonic quantum devices and circuits are built \cite{Kimble2008,RevModPhys.87.347,RevModPhys.87.1379,Haroche2020}.
Recently, cavity optomechanics has emerged in which the coupling between photonic and mechanical (phononic) modes allows coherent phonons to be controlled via photons, or vice versa \cite{Eichenfield2009,10.1038/nature08524,RevModPhys.86.1391,Guo2022,Yu2022}.
Such cavity optomechanics opens up entirely new perspectives for applications in ultra-precise sensing and meteorology, as well as building nano-opto-electro-mechanical devices with novel functionalities \cite{doi:10.1063/1.4896029,Midolo2018,doi:10.1515/nanoph-2021-0256}.

In particular, quantum emitters simultaneously coupled to cavity photonic and mechanical modes realize an interface between spin, photons and phonons as a hybrid quantum node.
However, the experimental exploration of such multi-modal systems remains a challenge.
For traditional emitters such as quantum dot and transition metal dichalcogenides, the coupling of electronic and mechanical degrees of freedom is weak, and thereby, typically studied in acoustic cavities \cite{Czerniuk2014,doi:10.1021/nl501413t,doi:10.1021/acs.nanolett.0c05089}.
In such system, the mechanical vibration is controlled by extrinsic driving, rather than the intrinsic light-matter coupling.
This limits the exploration of coupled mode physics and transduction between different degrees of freedom.

The light-matter coupling for traditional emitters can be described by the Jaynes-Cummings model \cite{PhysRevB.65.235311} where the electronic transition couples directly to the optical field without the participation of phonons.
In contrast, the charged boron vacancy $V_B^-$ in hexagonal boron nitride hBN is a recently explored quantum emitter \cite{2004.07968,Gottscholl2020,Stern2022,Gottscholl2021,doi:10.1021/acsphotonics.1c00320} that has a weak zero-phonon line, emission instead being dominated by phonon-induced processes \cite{Ivady2020,PhysRevB.102.144105,PhysRevLett.128.167401,doi:10.1021/acs.nanolett.2c00739}.
This is indicative of robust electron-phonon polarons that couple to the optical field, thereby intertwining electronic, phononic and photonic degrees of freedom.
By creating $V_B^-$ centers in an hBN nanobeam cavity \cite{PhysRevLett.128.237403,2204.04304}, we realize a novel emitter-optomechanical system for which $V_B^-$ related phonon polaritons couple electronic transitions, cavity photonic modes and cavity mechanical modes.

We investigate the emitter-optomechanical system using spatially resolved photoluminescence (PL) and Raman spectroscopy.
A pronounced spectral asymmetry is observed in the PL from the cavity photonic mode, only when the Q-factor is above a threshold of $\sim$10$^4$.
To explain this asymmetry, we construct a numerical model based on $V_B^-$ \cite{Ivady2020,PhysRevB.102.144105,PhysRevLett.128.167401,doi:10.1021/acs.nanolett.2c00739} for which the light-matter coupling is induced by the phonons.
Our model accounts well for the experiment, in contrast to expectations based on the Jaynes-Cummings model.
Furthermore, we demonstrate the strong interaction between phonon-induced $V_B^-$ emission and optomechanical cavity modes by spatially correlating luminescence in freely suspended structures \cite{D1NR07919K} and resonantly exciting the cavity photonic mode \cite{RevModPhys.86.1391}.
In position-dependent measurements, we observe anticrossings and phonon polaritons induced by the nanomechanical vibrations.
When subject to resonant excitation, the Raman signals arising from $V_B^-$ phonons exhibit an asymmetric enhancement for red and blue detunings, corresponding to cooling and heating of the cavity mechanical modes, respectively \cite{RevModPhys.86.1391}.
This reveals that the cavity mechanical modes induce the coupling between cavity photons and $V_B^-$ phonons.

% \section{\label{sec2}Results and Discussions}

\begin{figure*}
    \includegraphics[width=\linewidth]{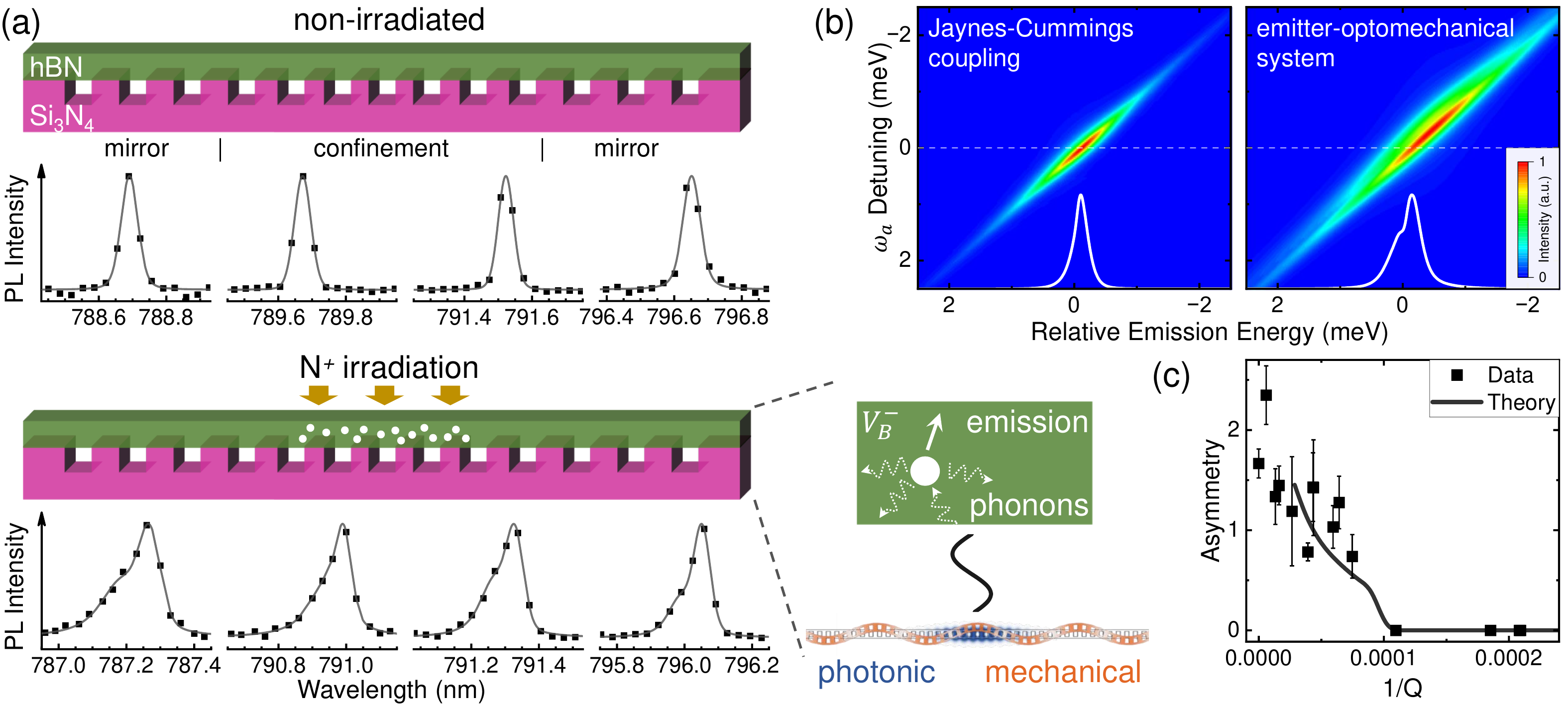}
    \caption{\label{f1}
        (a) Schematic of cavity structure for Sample-A.
        No asymmetry is observed in the lineshape of cavity peak when the hBN is not irradiated.
        In contrast, the irradiated cavities exhibit pronounced asymmetric lineshape arising from the emitter-optomechanical interaction depicted in the inset.
        (b) Calculated spectra of the cavity mode, in the cases of Jaynes-Cummings or phonon-induced light-matter coupling.
        White lines are the spectra at $\omega_a=\omega_X$ for examples.
        The phonon-induced coupling clearly arises the asymmetry of cavity emission lineshape.
        (c) The asymmetry has a threshold cavity Q-factor of 10000.
    }
\end{figure*}

The structure of our hBN/Si$_3$N$_4$ nanobeam cavity is schematically depicted in Fig.~\ref{f1}(a).
Fabrication procedures are presented in supplement Sec.~I A.
The confinement for nanophotonic modes is achieved by locally chirping the photonic crystal periodicity around the cavity center.
Besides the photonic modes, the cavity also supports nanomechanical modes, since the hBN/Si$_3$N$_4$ nanobeam is freely suspended but clamped at both ends \cite{2204.04304}.
Representative photonic and mechanical modes are depicted in Fig.~\ref{f1}(a) in blue and orange, respectively.
We investigate two samples in this work that differ with respect to the area over which $V_B^-$ centers are created.
In Sample-A, a 30 keV N$^+$ ion beam of $10^{13}$ $\mathrm{ions/cm^2}$ fluence (dose) creates $V_B^-$ centers only within the volume of photonic mode (Fig.~\ref{f1}(a) bottom).
The phonon-induced emission of $V_B^-$ \cite{Ivady2020,PhysRevB.102.144105,PhysRevLett.128.167401,doi:10.1021/acs.nanolett.2c00739} gives rise to a spectrally broad emission, even at low temperature \cite{2004.07968}, and coupling of local phonon modes to the optical transitions of $V_B^-$ center.
Thereby, phonons of the cavity mechanical modes may reasonably be expected to impact on the $V_B^-$ emission.
This cavity-$V_B^-$ system involves couplings between electronic transitions, $V_B^-$ phonons, cavity photons and cavity nanomechanical vibrations, as a multi-modal emitter-optomechanical system depicted schematically in Fig.~\ref{f1}(a).

\begin{figure*}
    \includegraphics[width=\linewidth]{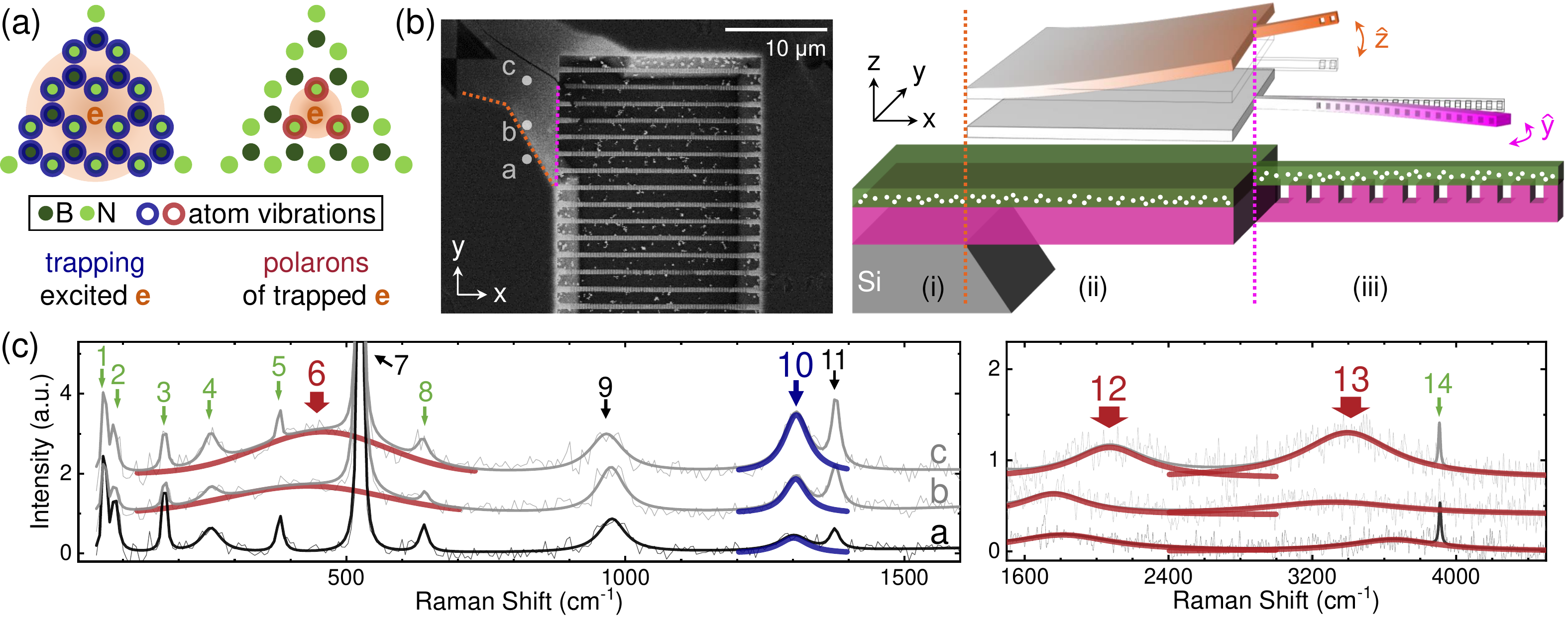}
    \caption{\label{f2}
        (a) Schematic of $V_B^-$-related phonon processes.
        Due to spatial overlap between electron wavefunction (brown shade) and atoms involved in the local phonon modes (blue and red rings), the extended phonon from the defect complex (blue rings) is expected to dominate the trapping process, and the highly localized phonons (red rings) dominate the polarons.
        (b) SEM image of Sample-B, including the large suspended membrane (upper left) and an array of nanobeam cavities (middle).
        The sample structure and corresponding mechanical modes are schematically shown at right.
        (c) Typical Raman spectra recorded from positions $a$-$c$ shown on (b).
        Phonon modes 1-14 are observed.
        Peak 10 is the extended phonon \cite{doi:10.1021/acs.chemmater.1c02849}.
        Broad peaks 6, 12 and 13 are highly localized phonons \cite{doi:10.1021/acs.chemmater.1c02849} and are the only ones exhibiting distinguishable position-dependent energy shift.
    }
\end{figure*}

We continue by comparing the emission spectral form observed from the cavities with and without irradiation.
All cavities are excited using a 532nm cw-laser having a spot size $\sim 1\ \mathrm{\mu m}$ and a power $\sim 120\ \mathrm{\mu W}$.
The cavity photonic mode is hence excited by the emission of $V_B^-$ centers in hBN (irradiated) and/or filtered light arising from other native luminescent defects in the hBN and underlying Si$_3$N$_4$ (non-irradiated).
The cavity mode exhibits high Q-factor $\sim 10^5$ limited only by the spectral resolution of our detection system \cite{PhysRevLett.128.237403}.
Remarkably, for the N$^+$ irradiated cavities we observe a strongly asymmetric lineshape of cavity emission that is broadened on the short wavelength (high energy) side.
Typical spectra recorded from four representative cavities without and with irradiation are presented in the upper and lower panels of Fig.~\ref{f1}(a), respectively.
We identify this asymmetry as arising from strong photon-phonon coupling \cite{RevModPhys.86.1391,10.1364/OE.27.030692,Guo2022}.
To quantitatively account for this spectral asymmetry, we construct a model in which a 2-level emitter simultaneously couples to cavity photons and local phonons.
The Hamiltonian of the multi-modal cavity QED system can then be written as
\begin{eqnarray}
    \label{eqH1}
    \textit{H} &=& \hbar\omega_{X}\sigma_{X,X} + \hbar\omega_{a}a^{+}a + \hbar\omega_{q}q^{+}q \nonumber \\
    &\;& + \hbar \lambda_{e\text{-}q}\sigma_{X,X}\left(q+q^{+}\right) + \hbar \lambda_{p\text{-}q}a^{+}a\left(q+q^{+}\right) \nonumber \\
    &\;& + \textit{H}_{e\text{-}p} \nonumber
\end{eqnarray}
including the 6 terms corresponding to the emitter, photon, phonon, emitter-phonon coupling with a strength $\lambda_{e\text{-}q}$, photon-phonon coupling with a strength $\lambda_{p\text{-}q}$, and emitter-photon coupling, respectively \cite{PhysRevB.65.235311}. $\sigma_{A,B}$ ($A,B\in\{X,G\}$) is the Dirac operator for the emitter with the ground ($G$) and excited ($X$) state, $a^{+}$/$a$ are the ladder operators for cavity photons, and $q^{+}$/$q$ are the ladder operators for phonons.
Since the $V_B^-$ emission is dominated by phonon-induced processes \cite{Ivady2020,PhysRevB.102.144105,PhysRevLett.128.167401,doi:10.1021/acs.nanolett.2c00739}, we introduce an effective phonon-induced emitter-photon coupling
\begin{eqnarray}
    \label{eqH2b}
    \textit{H}_{e\text{-}p} &=& \hbar g\left(\sigma_{G,X}\left(q^{+}\right)^k a^{+} + \sigma_{X,G}q^k a\right) \nonumber
\end{eqnarray}
where $g$ is the coupling strength and $k$ is the number of phonons involved the coupling.
$k=0$ corresponds to the conventional Jaynes-Cummings model applied for usual emitters dominated by zero-phonon line emission, where the phonons only introduce vibronic sublevels to \textit{mediate} the emitter-photon coupling, but do not directly participate.
In contrast, $\textit{H}_{e\text{-}p}$ with $k \geq 1$ represents an entirely different but coherent interface of emitter, photon and phonon, in which the phonon is necessary to \textit{induce} the emitter-photon coupling.

We calculated the spectral form of the emission from the cavity photonic mode by solving the master equation (for details see supplement Sec.~I D).
Typical spectra in the case of $k=0$ and $k=4$ are presented in Fig.~\ref{f1}(b), as the energy detuning between 2-level emitter and cavity photonic mode is varied.
The white curves show the spectral form at resonance $\omega_a=\omega_X$.
The asymmetry in the emission lineshape is suppressed for the Jaynes-Cummings coupling ($k=0$) but becomes pronounced in the emitter-optomechanical interaction ($k=4$).
Despite the model being rather simple, it captures the essential physics underpinning our experimental observations.
As presented in Fig.~\ref{f1}(a), we do not observe any asymmetry in any of our control experiments where the hBN was not irradiated with N$^+$.
This is the central result of this work, showing that the phonons involved in the photon-phonon coupling are not thermally excited, but rather arise during the light-matter coupling of $V_B^-$ centers.
As such, they give rise to a novel emitter-induced optomechanical coupling.

We note that the asymmetric cavity lineshape is only observed for high-Q irradiated cavities.
We use bi-peak fitting to quantify the asymmetry via the ratio between the intensity of two peaks (for details see supplement Sec.~I C) and present
the results in Fig.~\ref{f1}(c).
A threshold Q-factor of $\sim 10^4$ is clearly observed, meaning that the $V_B^-$-induced photon-phonon coupling rate needs to exceed the system decay rate to observe the asymmetric lineshape.
This threshold behavior is well reproduced in the theoretical calculation, as presented by the solid line in Fig.~\ref{f1}(c) for the case $k=4$.

Generally, as schematically depicted in Fig.~\ref{f2}(a), two different phonon processes are involved in the $V_B^-$ emission process:
Firstly, the negatively charged electronic state $V_B^-$ is generated by trapping an excited electron that relaxes via phonon-assisted processes \cite{Shan2014,PhysRevB.56.10435}.
Secondly, the excited electron decays again as the polaron with localized phonons to produce a broad phonon sideband in the PL spectrum.
The electron-phonon spatial overlap \cite{doi:10.1021/acsnano.8b04803,Merkl2021} depicted schematically in Fig.~\ref{f2}(a) indicates that, extended phonons involving the vibration of multiple atoms around $V_B^-$ (blue rings) govern the trapping photo-physics, whilst more localized phonons primarily involving neighboring atoms (red rings) dominate the polarons.
J. Li et al. \cite{doi:10.1021/acs.chemmater.1c02849} recently reported Raman studies of $V_B^-$ and unveiled these extended and highly localized phonons via boron isotope characterization.

\begin{figure}
    \includegraphics[width=\linewidth]{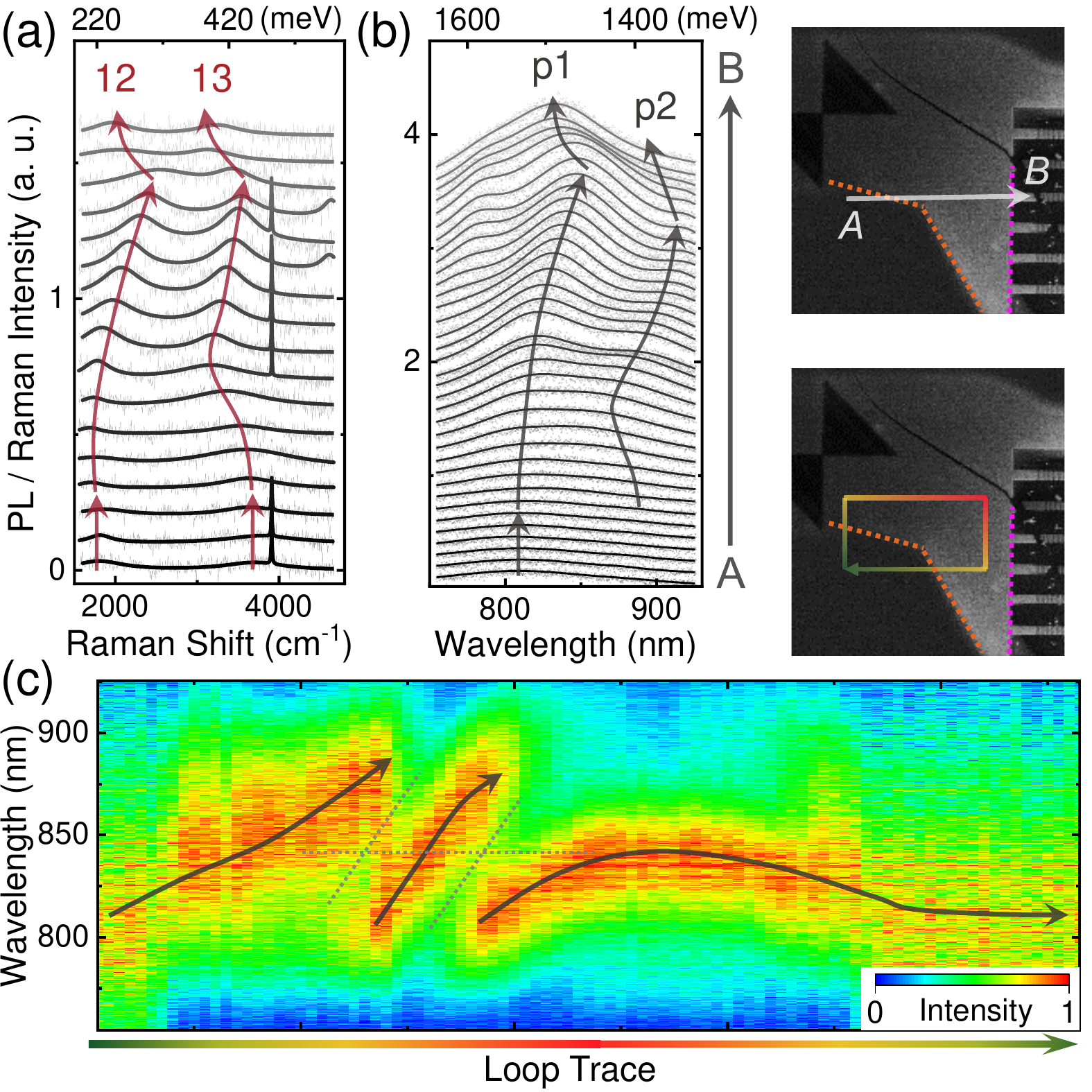}
    \caption{\label{f3}
        Spatially resolved Raman (a) and PL (b) spectra recorded from positions along $A\rightarrow B$ in the inset.
        (c) Map of normalized PL spectra extracted from a loop trace shown in the inset.
        Anticrossing is clearly demonstrated, otherwise, the PL peak cannot shift back after the loop.
    }
\end{figure}

To confirm this picture for $V_B^-$ emission in Fig.~\ref{f2}(a) and the interaction with mechanical modes in nanocavities, we recorded spatially resolved PL and Raman spectroscopy from another sample (Sample-B) that is homogeneously irradiated by 30 keV N$^+$ ions with a fluence (dose) of $10^{14}\ \mathrm{ions/cm^2}$ \cite{doi:10.1021/acs.nanolett.2c00739,doi:10.1021/acsphotonics.0c00614,doi:10.1021/acsomega.1c04564} as depicted in Fig.~\ref{f2}(b).
Sample-B contains three different regions of interest:
(i) the non-underetched region with hBN on a planar Si$_3$N$_4$/Si substrate corresponding to the dark surrounding region in the SEM image,
(ii) the large suspended membrane with hBN/Si$_3$N$_4$ layers suspended on underetched Si corresponding to the large bright region at left top,
and (iii) suspended nanobeam cavities.
Typical Raman spectra recorded from three different positions labelled $a$-$c$ in Fig.~\ref{f2}(b) are presented in Fig.~\ref{f2}(c).
The observed Raman features are labelled 1-14.
Peaks 7 and 9 at 524 and 975 $\mathrm{cm^{-1}}$ are the bulk Si$_3$N$_4$ phonons \cite{IATSUNSKYI20151650}.
Peak 11 at 1377 $\mathrm{cm^{-1}}$ is the bulk hBN phonon $E_{2g}$ \cite{doi:10.1021/acs.chemmater.1c02849}.
The other peaks are not observed in Si$_3$N$_4$ nor non-irradiated hBN (see supplement Sec.~II B).
Hence, they are indentified as being related to the $V_B^-$ centers.
Peak 6 around 450 $\mathrm{cm^{-1}}$ (red curve) is the highly localized phonon, and peak 10 at 1306 $\mathrm{cm^{-1}}$ (blue) is the extended phonon depicted in the context of Fig.~\ref{f2}(a), which have been previously identified \cite{doi:10.1021/acs.chemmater.1c02849}.
We observe two other broad Raman features at higher energies: peak 12 around 1800 $\mathrm{cm^{-1}}$ (quadruple of peak 6) and peak 13 around 3600 $\mathrm{cm^{-1}}$ (double of peak 12).
Broader Raman linewidth typically implies stronger spatial confinement of the phonon mode \cite{PhysRevLett.113.175501, doi.org/10.1002/jrs.1684,doi.org/10.1002/jrs.5815}.
In addition, peaks 6, 12 and 13 are the only $V_B^-$ phonons exhibiting distinguishable energy shifts when moving the laser spot e.g., between the three positions in Fig.~\ref{f2}(c).
These common features indicate that peaks 12 and 13 stem from multi-phonon states of peak 6, and are also highly localized at the $V_B^-$ center.
Peaks 1-5, 8 and 14 are discussed in supplement Sec.~III B, due to they exhibit little correlation to the $V_B^-$ emission.

In Fig.~\ref{f3}(a)(b), we present Raman and PL spectra of $V_B^-$ recorded from positions along the line $A\rightarrow B$ illustrated in the inset.
This line includes the three regions (i)-(iii) depicted in Fig.~\ref{f2}(b).
Position-dependent energy shifts of the Raman and PL peaks in the three regions are marked with the arrows.
Their generic shifts exhibit remarkable correlations i.e., the shifts of Raman peak 12 and 13 in Fig.~\ref{f3}(a) track the shifts of PL peak p1 and p2 in \ref{f3}(b).
This observation strongly suggests that PL peak p1 and p2 are phonon sidebands arising from the highly localized phonons Raman peak 12 and 13.
Moreover, we observe that the position-dependent shifts of PL peaks and $V_B^-$ phonons exhibit pronounced anticrossings.
Fig. \ref{f3}(c) shows PL spectra recorded from various positions along a closed clockwise loop on the sample illustrated in the inset.
As depicted by the dashed lines in Fig.~\ref{f3}(c), two anticrossings are observed at which one spectral feature disappears and is replaced by another.
This combined with the identity of spectra at the beginning and end of loop provide strong evidence for the phonon polaritons induced by the position-dependent mechanical modes.
Similar feature and anticrossings between highly localized phonons are also observed from the spatially resolved Raman spectra (see supplement Sec.~II E), further supporting this conclusion.

\begin{figure}
    \includegraphics[width=\linewidth]{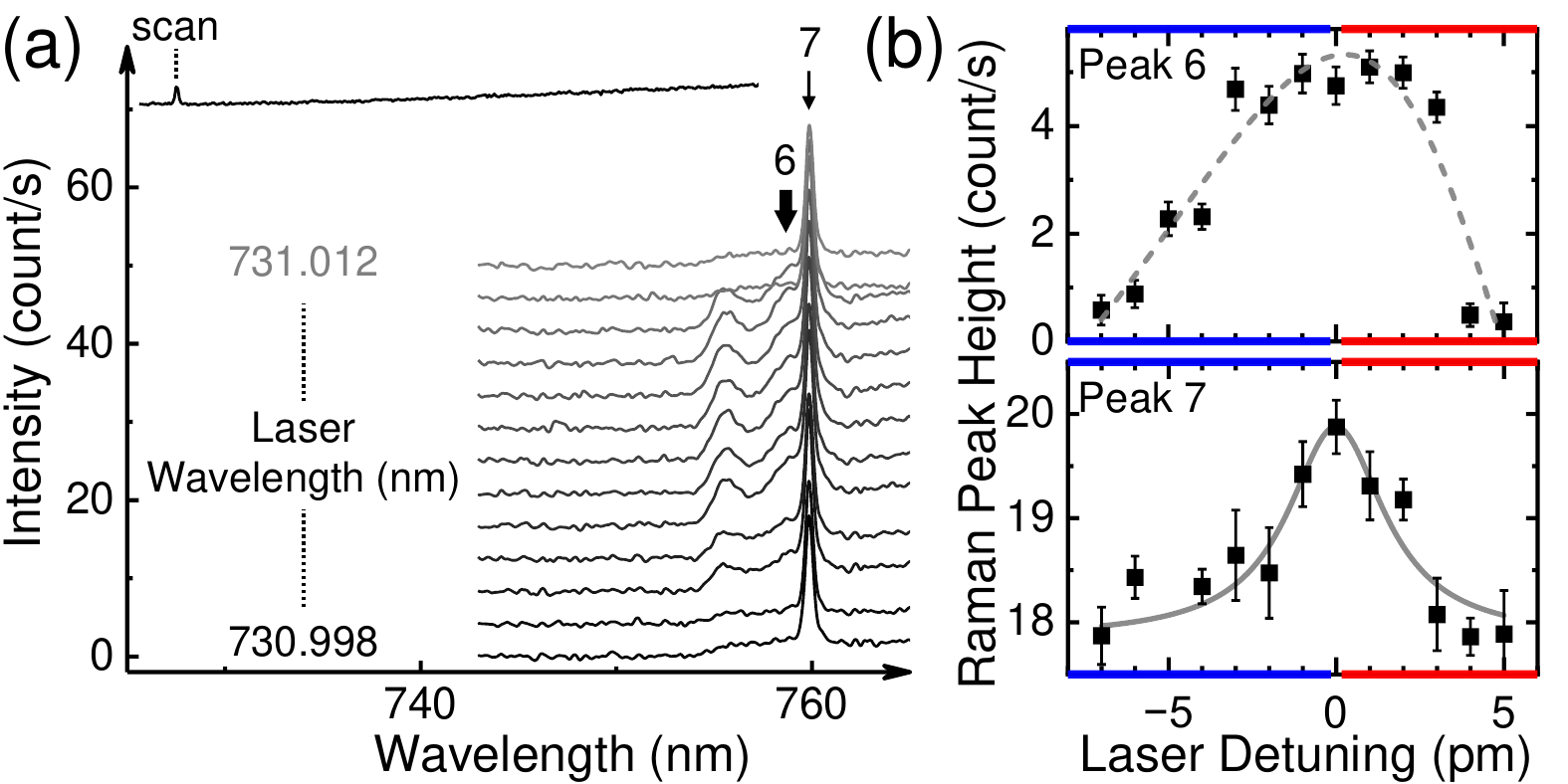}
    \caption{\label{f4}
        (a) Resonant excitation of a cavity photonic mode that is denoted in the top spectrum excited by 532nm laser.
        (b) The enhancement of $V_B^-$ phonon peak 6 is asymmetric: strong at blue (heating) and weak at red (cooling) detunings for cavity mechanical modes.
        In contrast, the enhancement of Si$_3$N$_4$ phonon peak 7 is nearly symmetric.
    }
\end{figure}

Finally, we explore the emitter-optomechanical interaction by controlling the optomechanical cavity modes via resonant excitation of the cavity itself.
Hereby, the spectrum is detected as a narrowband ($\leq 1$MHz) cw-laser is tuned from 730.998 and 731.012 nm encompassing an ultra-high-Q cavity photonic mode at $\sim$ 731 nm.
Typical results obtained from such resonant excitation are presented in Fig.~\ref{f4}(b) which shows the intensity of Raman peak 6 and 7 as a function of laser-cavity detuning.
The population of photons in the cavity is enhanced for both red and blue detunings.
The enhanced cavity pho\textit{tons} couple to the pho\textit{nons} of Si$_3$N$_4$ and $V_B^-$ thus enhance the corresponding Raman signals \cite{doi:10.1080/05704928.2019.1661850}.
In contrast, the population of cavity mechanical phonons is reduced at red detunings (cooling) but enhanced at blue detunings (heating) \cite{RevModPhys.86.1391}.
As presented in Fig.~\ref{f4}(b) upper panel, the enhancement of peak 6 arising from $V_B^-$ phonon is pronounced, varying by $>10\times$ as the laser is tuned through the cavity photonic mode.
Moreover, this enhancement is asymmetric with respect to detuning, being weaker with the cooling of cavity vibrations (red detuning) but stronger with the heating (blue detuning).
This observation reveals that the cavity mechanical modes induce the coupling between cavity photons and $V_B^-$ phonons \cite{2204.04304}.
In contrast, the enhancement of Si$_3$N$_4$ phonon peak 7 is much weaker ($10\%$) and nearly symmetric, indicating that the cavity photons couple to Si$_3$N$_4$ phonon directly without the participation of cavity vibrations.

% \section{\label{sec3}Conclusions}

In summary, we report on the interaction between $V_B^-$ centers and optomechanical modes in a hBN nanocavity.
PL and Raman data demonstrate the $V_B^-$-induced cavity optomechanical coupling and the control of $V_B^-$ transitions and polarons through cavity optomechanics.
This emitter-optomechanical interaction stems from the ultra-high Q-factor of our nanocavity and the phonon-induced light-matter coupling.
These results reveal that the coupling between multiple degrees of freedom arise from the intrinsic phonon processes in the system.
The phononic features are strong since hBN is an atomically thin layered material that is highly sensitive to local deformation \cite{1911.08072,Autore2018,doi:10.1021/acsphotonics.9b01094,PhysRevLett.126.227401}.
Our work extends cavity QED to the regime where electronic systems are simultaneously coupled to both phononic and photonic degrees of freedom.
We believe that such hybrid interface in nanosystem will open up interesting perspectives for quantum sensing and meteorology, as well as quantum transduction.

\begin{acknowledgments}
    All authors gratefully acknowledge the German Science Foundation (DFG) for financial support via grants FI 947/8-1, DI 2013/5-1, AS 310/9-1 and SPP-2244, as well as the clusters of excellence MCQST (EXS-2111) and e-conversion (EXS-2089).
    J. J. F. gratefully acknowledges the state of Bavaria via the One Munich Strategy and Munich Quantum Valley.
    C. Q. and V. V. gratefully acknowledge the Alexander v. Humboldt foundation for financial support in the framework of their fellowship programme.
    Support by the Ion Beam Center (IBC) at HZDR is gratefully acknowledged.

\end{acknowledgments}

\end{document}

\title{Supplementary Information for Emitter-Optomechanical Interaction in Ultra-High-Q hBN Nanocavities}

\author{Chenjiang Qian}
\email{chenjiang.qian@wsi.tum.de}
\author{Viviana Villafañe}
\author{Martin Schalk}
\affiliation{Walter Schottky Institut and Physik Department, Technische Universit{\" a}t M{\" u}nchen, Am Coulombwall 4, 85748 Garching, Germany}
\author{Georgy V. Astakhov}
\author{Ulrich Kentsch}
\author{Manfred Helm}
\affiliation{Helmholtz-Zentrum Dresden-Rossendorf, Institute of Ion Beam Physics and Materials Research, 01328 Dresden, Germany}
\author{Pedro Soubelet}
\author{Andreas V. Stier}
\affiliation{Walter Schottky Institut and Physik Department, Technische Universit{\" a}t M{\" u}nchen, Am Coulombwall 4, 85748 Garching, Germany}
\author{Jonathan J. Finley}
\email{finley@wsi.tum.de}
\affiliation{Walter Schottky Institut and Physik Department, Technische Universit{\" a}t M{\" u}nchen, Am Coulombwall 4, 85748 Garching, Germany}
%\date{\today}

\maketitle

\tableofcontents

\section{\label{fms} Methods}

\subsection{\label{sfa} Sample Fabrication}

\begin{figure}
    \includegraphics[width=\linewidth]{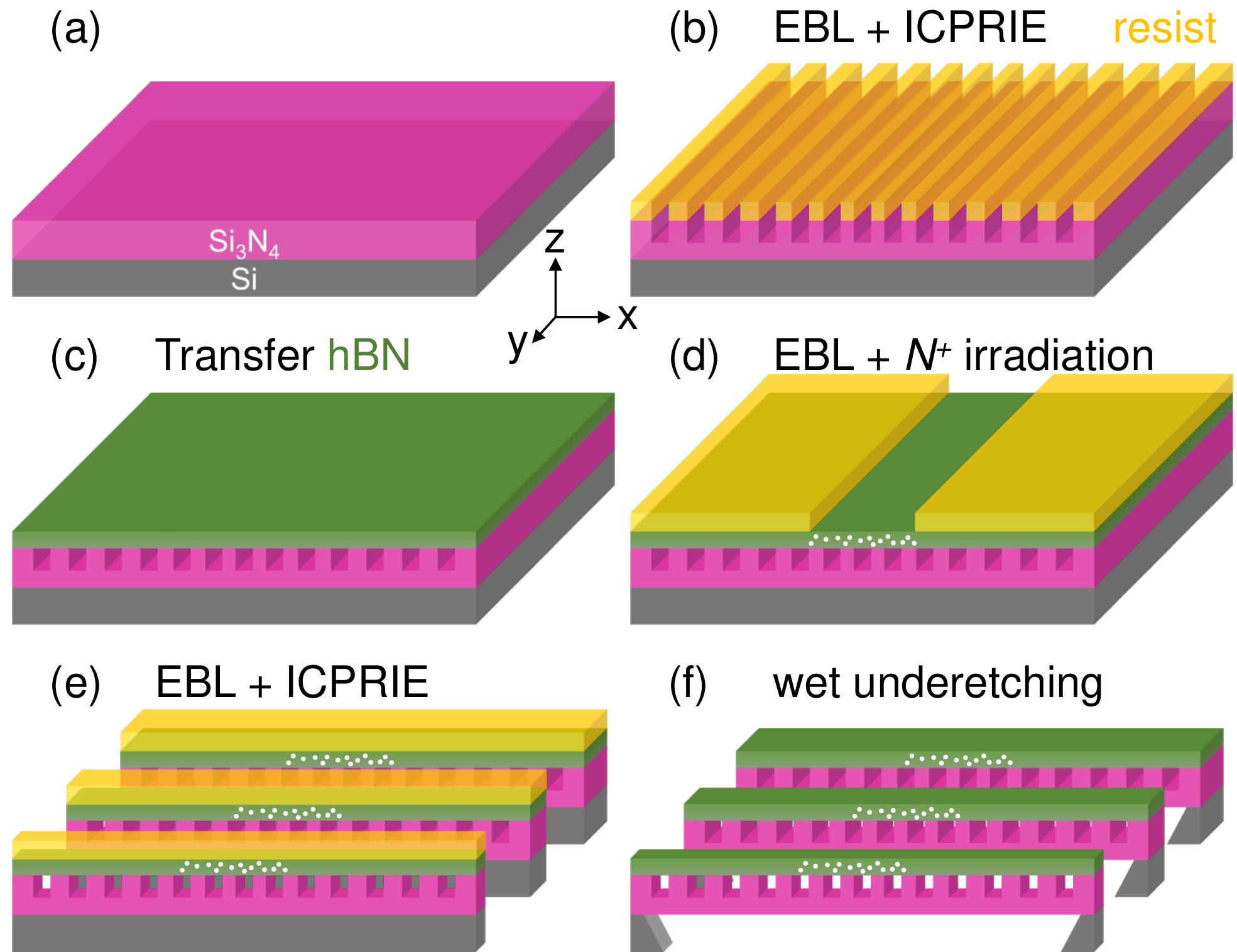}
    \caption{\label{fab}
        Fabrication procedures.
        (a) Si$_3$N$_4$/Si substrate.
        (b) EBL and ICPRIE to fabricate periodic nanotrenches.
        (c) Transfer the hBN flake on top of the nanotrenches.
        (d) Generate $V_B^-$ site-selectively.
        (e) EBL and ICPRIE to divide the nanobeams.
        (f) Wet underetching to remove the bottom Si.
    }
\end{figure}

The hBN/Si$_3$N$_4$ hybrid nanobeam cavity is fabricated through e-beam lithography (EBL), inductively coupled plasma reactive ion etching (ICPRIE), viscoelastic dry transfer method \cite{Pizzocchero2016} and wet underetching.
Fabrication procedures of Sample-A are schematically described in SFig.~\ref{fab}.
Firstly, we prepare a Si substrate with 200nm Si$_3$N$_4$ grown by low pressure chemical vapor deposition (LPCVD) on top, as shown in SFig.~\ref{fab}(a).
Then we use EBL and ICPRIE to etch periodic nanotrenches in Si$_3$N$_4$ as shown in SFig.~\ref{fab}(b).
After removing the residual resist, we transfer the hBN flake with a thickness $\sim 100\ \mathrm{nm}$ on top of the nanotrenches, as shown in SFig.~\ref{fab}(c).
We coat the hBN flake by the resist mask and use EBL to pattern a window at the cavity center, followed by the ion irradiation to create $V_B^-$ centers through the window, as shown in SFig.~\ref{fab}(d).
In this step the resist has the thickness $>600\ \mathrm{nm}$ to protect the under lying hBN \cite{doi:10.1021/acsomega.1c04564}.
Finally, we use EBL and ICPRIE to divide the nanobeams, as shown in SFig.~\ref{fab}(e), followed by a wet underetching to remove the bottom Si, as shown in SFig.~\ref{fab}(f).
The fabrication procedures of Sample-B are generally the same.
The only difference is that the $V_B^-$ centers in Sample-B are not created site-selectively, but rather homogeneously in the whole hBN flake (without any resist mask) after all other fabrication procedures.
Detailed parameters used in the fabrication and calculation results of cavity modes have been reported in our previous works \cite{PhysRevLett.128.237403,doi:10.1021/acs.nanolett.2c00739,2204.04304}.

\subsection{\label{mss} Measurement Setup}

Photoluminescence (PL) and Raman spectroscopy in this work are recorded using a confocal micro-PL setup.
The objective with a magnification of 100 $\times$ and a NA of 0.75 is used to focus the laser into a spot with the size $\sim1\ \mathrm{\mu m^2}$ on the sample.
The resonant excitation of cavity photons (Fig.~4) is carried out with a tunable narrow band CW-laser.
Other experiments are carried out with a 532nm CW-laser.
The samples are mounted on a three-dimensional xyz nanopositioner for spatially resolved spectroscopy.
The PL and Raman spectra of $V_B^-$ centers (Fig.~2-4) are collected by a matrix array Si CCD detector in a spectrometer, with a focal length of 0.55 m and the grating of 300 grooves per mm.
The PL spectra of cavity photons (Fig.~1) are collected by the same detector and spectrometer but with the grating of 1200 grooves per mm.

\begin{figure}
    \includegraphics[width=\linewidth]{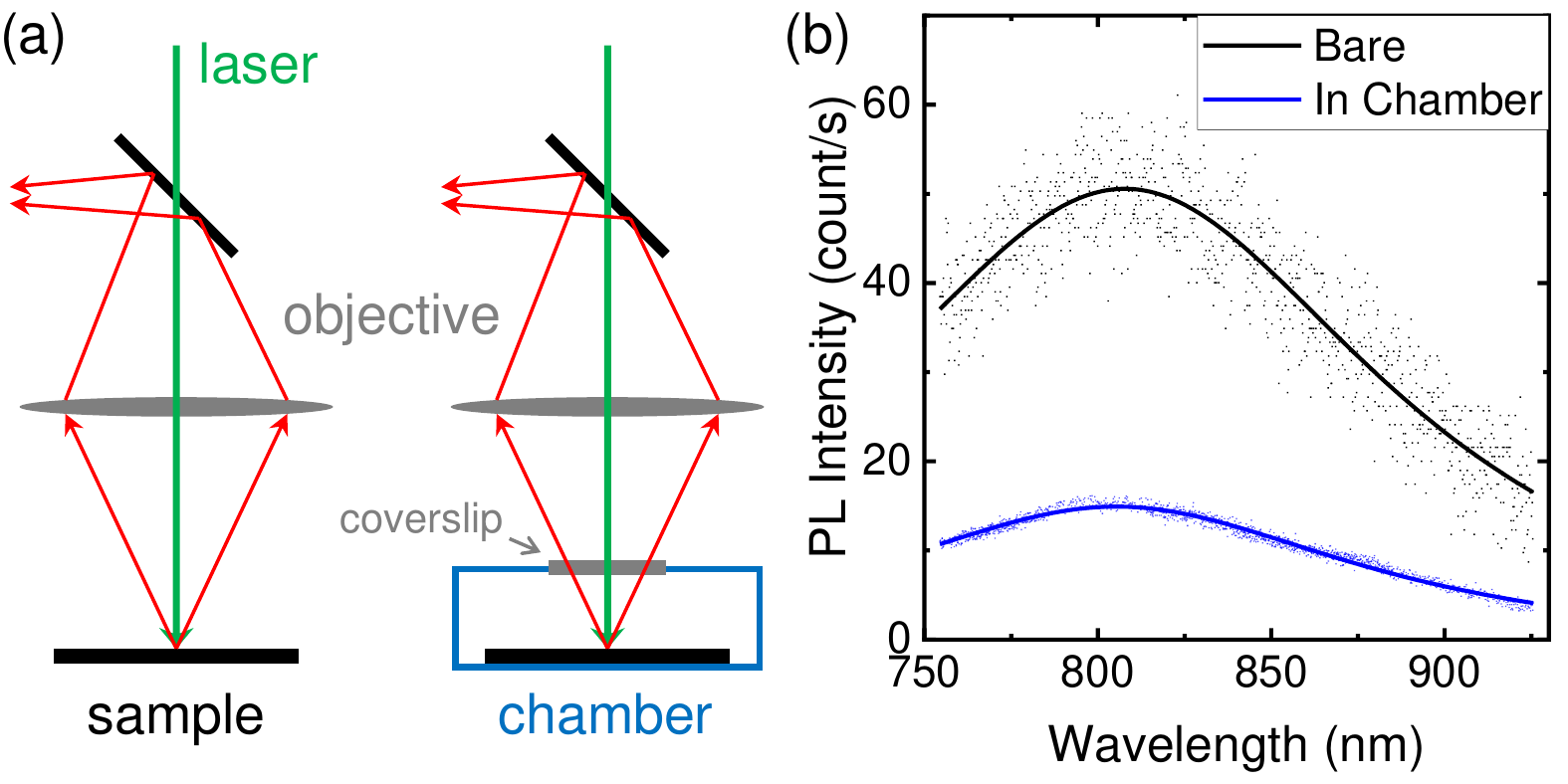}
    \caption{\label{wow}
        (a) Schematic of the confocal micro-PL setup, with the sample (left) bare in the air and (right) in the chamber.
        (b) Comparison of PL spectra of $V_B^-$ in bare (black) and in the chamber (blue), recorded at the same position with the same laser power of $264\ \mathrm{\mu W}$.
    }
\end{figure}

In this work, the sample is measured at room temperature in ambient conditions, since we found that the glass coverslip of our cryostat chamber suppressed the intensity and deformed the spectrum, particularly at long wavelength over 800 nm.
The confocal setup is schematically depicted in SFig.~\ref{wow}(a).
For a high NA objective, the coverslip was found to degrade the aplanatic and sine conditions, thereby introducing spherical and chromatic aberrations \cite{Diel2020,doi:10.2144/02334bi01}.
As a result, we observe the suppression of signals when the sample is measured in the chamber e.g., the comparison presented in SFig.~\ref{wow}(b).
Moreover, we find this suppression increases with the photon wavelength, by $\sim 25 \%$ from 800 to 920 nm. 
The lineshape of broad $V_B^-$ emission will be distorted by the non-uniform suppression.
Therefore, we measure the sample bare in the air to collect unperturbed signals and avoid potential errors.

\subsection{\label{fitt} Fitting Methods}

\begin{figure}
    \includegraphics[width=\linewidth]{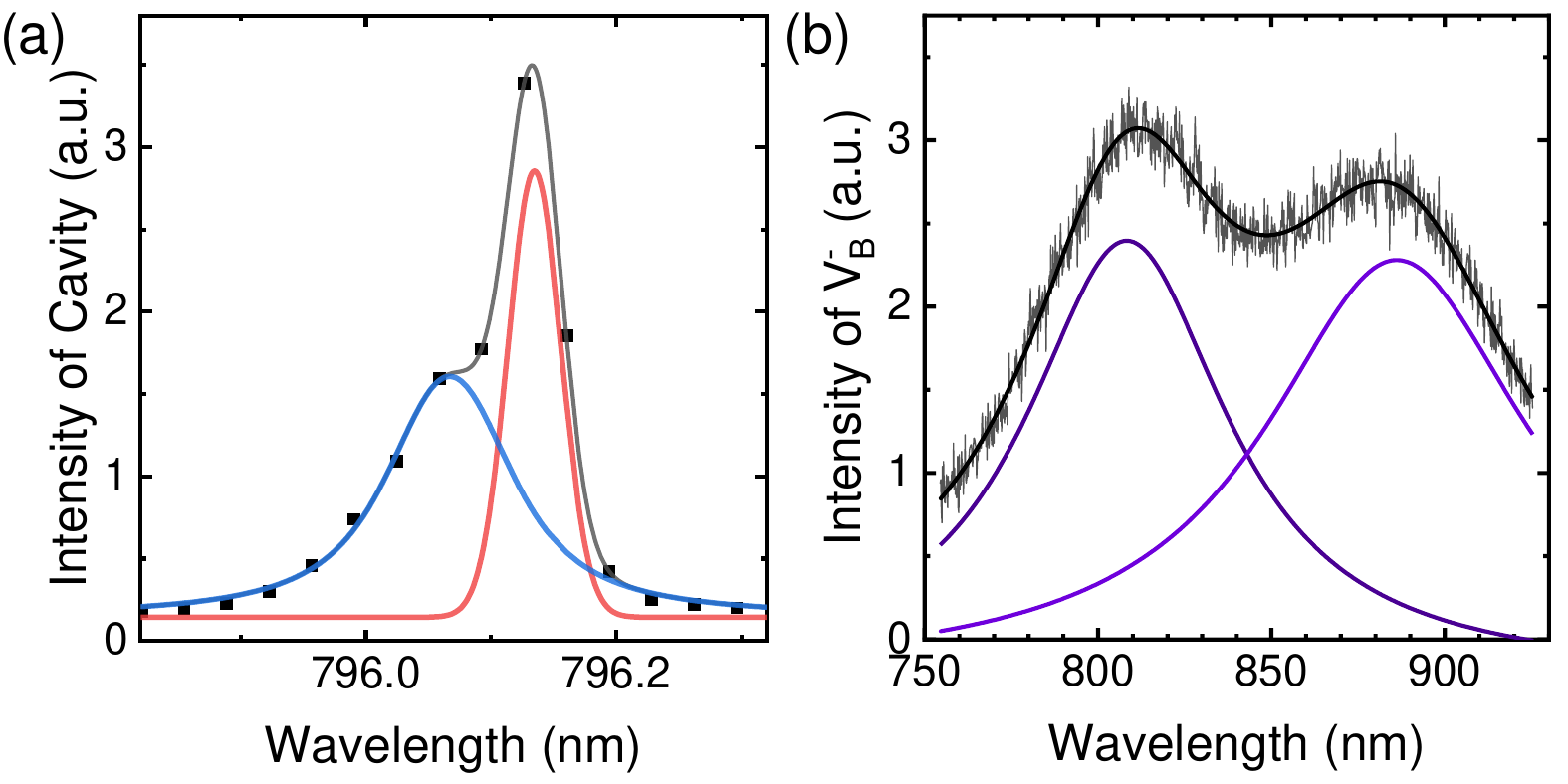}
    \caption{\label{plf}
        Fitting of (a) cavity photonic mode and (b) $V_B^-$ emission for examples.
    }
\end{figure}

The raw PL spectrum is broadened by our detection system, which generally follows the convolution to a Gaussian window function \cite{JIMENEZMIER1994741,Nomura2010}.
We use two Voigt peaks \cite{Nomura2010}, which is the convolution of Lorentz and Gaussian peak, to fit the PL peak of cavity photons as presented in SFig.~\ref{plf}(a).
The Gaussian width of Voigt peak is fixed at $\Delta_G=50\ \mathrm{pm}$.
The real cavity Q-factor after deconvolution is thereby calculated from the Lorentz width $\Delta_L$ of the major peak (red) from the fitting.
In addition, we use the relative intensity of the polaron peak (blue) to the major peak (red) as the quantitation of asymmetry presented in Fig.~1(c).
For the broad $V_B^-$ emission, the linewidth is much broader than the broadening of detection system.
Thus, we directly use Lorentz peaks in the fitting of $V_B^-$ emission peaks such as the example presented in SFig.~\ref{plf}(b).

\begin{figure}
    \includegraphics[width=\linewidth]{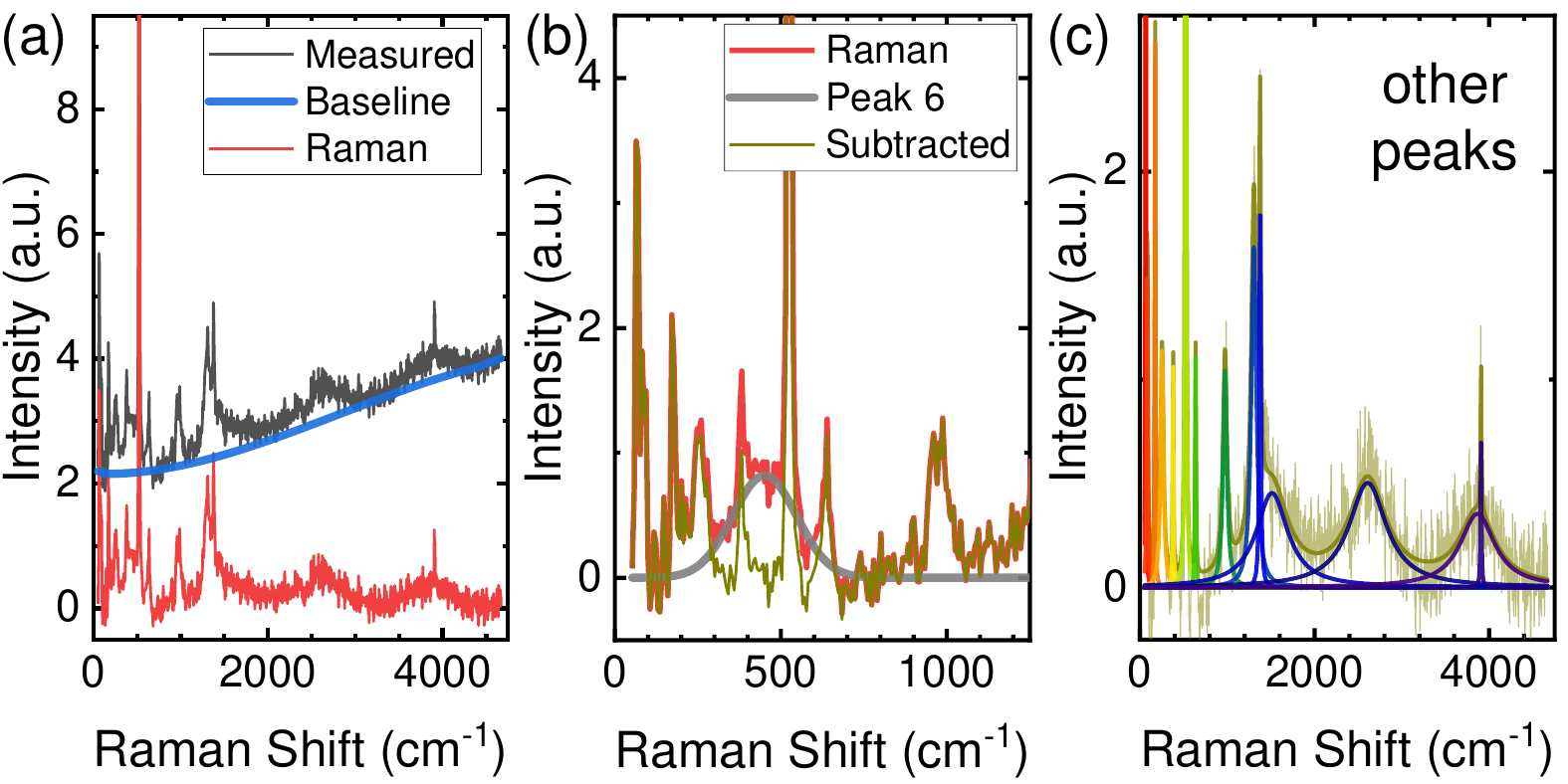}
    \caption{\label{raf}
        (a) Raman spectrum (red) is extracted by subtracting the measured spectrum (gray) with the emission baseline (blue).
        (b) Peak 6 overlaps with many other peaks.
        Thereby, we first extract and subtract peak 6 as a baseline (gray) from the Raman spectrum (red).
        (c) Other Raman peaks (color lines) are then extracted by the multi Lorentz fitting.
    }
\end{figure}

The recorded spectra around the wavelength of excitation laser, such as the gray line in SFig.~\ref{raf}(a), consist of Raman and PL signals superimposed.
As a common method, we first draw the emission baseline based on the peak foots of Raman signals, and then extract the Raman spectrum by subtracting the emission baseline from the recorded data, such as the example in SFig.~\ref{raf}(a).
In the Raman spectra, the broad peak 6 of highly localized phonon overlaps with many other peaks.
Thereby, it is not easy to fit all overlapped peaks together.
To separate these peaks, we first extract peak 6 by a Gaussian base peak as shown in SFig.~\ref{raf}(b).
Then other peaks are fitted by multi Lorentz peaks as shown in SFig.~\ref{raf}(c).
This fitting method might introduce a bit inaccuracy of peak 6.
Nonetheless, as revealed by our work, the $V_B^-$ emission wavelengths mainly depend on multi-phonon states of highly localized phonons represented by peak 12 and 13.
Thus, even if some minor inaccuracy exists in the fitting of peak 6, our analysis will not be affected.

\subsection{\label{meq} Theoretical Calculations}

Since specific dynamic and parameters of the phonon-induced processes during the $V_B^-$ emission are not fully known yet \cite{Ivady2020,PhysRevB.102.144105,PhysRevLett.128.167401}, we consider a brief system consisting of a single emitter described by the ground state $G$ and the excited state $X$ with the energy $\omega_X$ and decay rate $\gamma_X$, a single cavity photonic mode with the energy $\omega_a$ and decay rate $\gamma_a$, and a single phonon mode with the energy $\omega_q$ and decay rate $\gamma_q$.
The Hamiltonian of the cavity QED system is thereby wrote as
\begin{eqnarray}
    \label{eqH1}
    \textit{H} &=& \textit{H}_e + \textit{H}_p + \textit{H}_q + \textit{H}_{e\text{-}q} + \textit{H}_{p\text{-}q} + \textit{H}_{e\text{-}p}
\end{eqnarray}
involving the 6 terms correspond to emitter ($\textit{H}_e$), photon ($\textit{H}_p$), phonon ($\textit{H}_q$), emitter-phonon coupling ($\textit{H}_{e\text{-}q}$), photon-phonon coupling ($\textit{H}_{p\text{-}q}$), emitter-photon coupling ($\textit{H}_{e\text{-}p}$), respectively.
The first three terms are clearly defined as
\begin{eqnarray}
    \label{eqH1b}
    \textit{H}_e &=& \hbar\omega_{X}\sigma_{X,X},\  \nonumber \\
    \textit{H}_p &=& \hbar\omega_{a}a^{+}a \\
    \textit{H}_q &=& \hbar\omega_{q}q^{+}q \nonumber
\end{eqnarray}
where $\sigma_{A,B}=|A \rangle \langle B|$ ($A,B\in\{X,G\}$) is the Dirac operator for the emitter, $a^{+}$/$a$ are the ladder operators for photons in the cavity photonic mode, and $q^{+}$/$q$ are the ladder operators for phonons.
For an atomic $V_B^-$ defect, the oscillatory motion of confined electron is expected to be much faster than the phonon frequency (adiabatic
approximation) \cite{PhysRevB.65.235311}, thereby, the two phonon coupling terms are wrote as
\begin{eqnarray}
    \label{eqH1c}
    \textit{H}_{e\text{-}q} &=& \hbar \lambda_{e\text{-}q}\sigma_{X,X}\left(q+q^{+}\right) \nonumber \\
    \textit{H}_{p\text{-}q} &=& \hbar \lambda_{p\text{-}q}a^{+}a\left(q+q^{+}\right) 
\end{eqnarray}
where $\lambda_{e\text{-}q}$ and $\lambda_{p\text{-}q}$ are the strength for emitter-phonon and photon-phonon coupling, respectively.
The emitter-photon coupling with a strength $g$ is usually described by the Jaynes-Cummings model as 
\begin{eqnarray}
    \label{eqH2a}
    \textit{H}_{e\text{-}p} &=& \hbar g\left(\sigma_{G,X}a^{+} + \sigma_{X,G}a\right)
\end{eqnarray}
which means the $X \rightarrow G$ transition emit a photon to the cavity photonic mode, and a photon in the cavity photonic mode excite the emitter as $G \rightarrow X$.
However, the $V_B^-$ emission is dominated by the phonon-induced processes.
Hereby we introduce an effective phonon-induced emitter-photon coupling, e.g.,
\begin{eqnarray}
    \label{eqH2b}
    \textit{H}_{e\text{-}p} &=& \hbar g\left(\sigma_{G,X}q^{+}a^{+} + \sigma_{X,G}qa\right)
\end{eqnarray}
for one-phonon-induced emitter-photon coupling and 
\begin{eqnarray}
    \label{eqH2c}
    \textit{H}_{e\text{-}p} &=& \hbar g\left(\sigma_{G,X}\left(q^{+}\right)^k a^{+} + \sigma_{X,G}q^k a\right)
\end{eqnarray}
for $k$-phonon-induced emitter-photon coupling.

\begin{figure}
    \includegraphics[width=\linewidth]{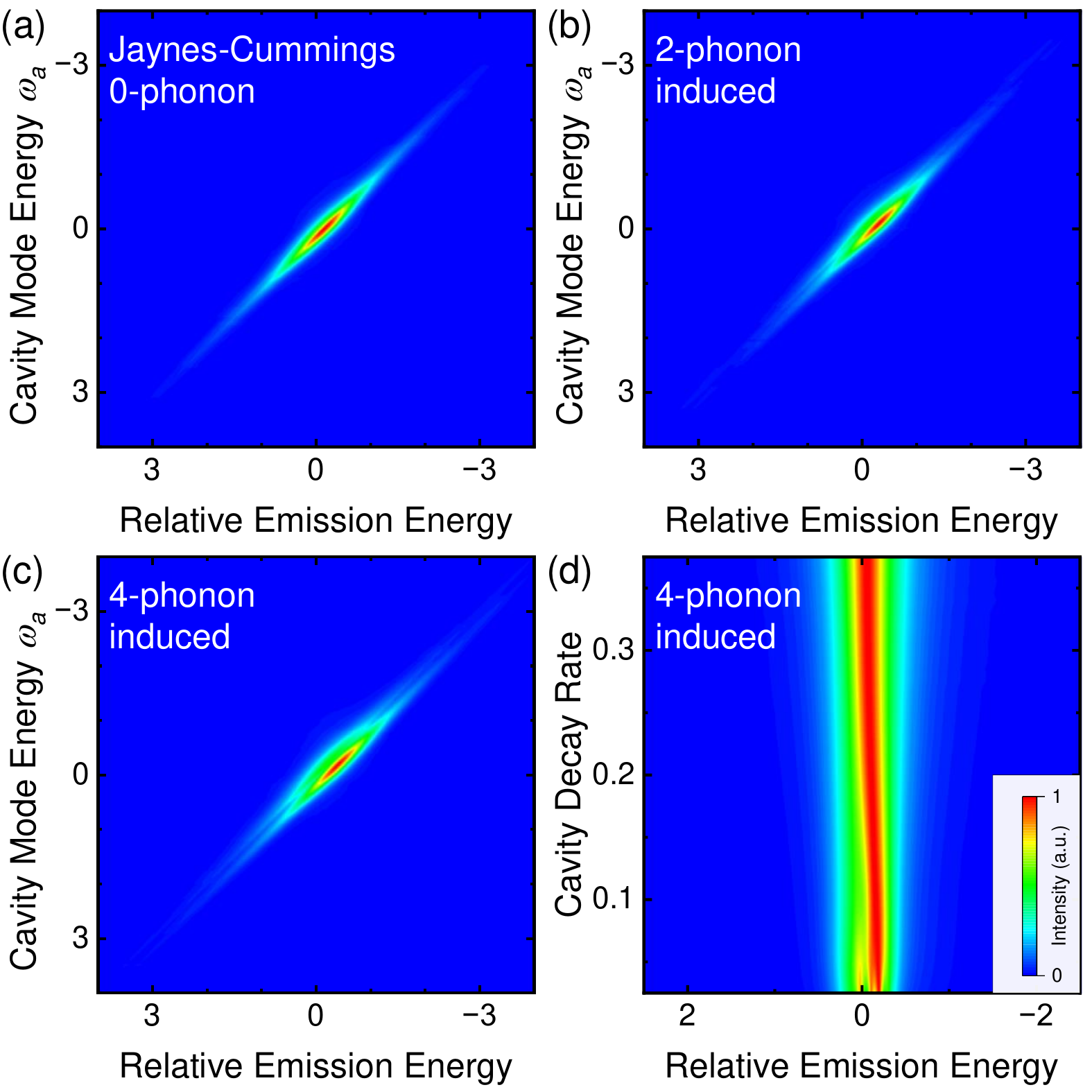}
    \caption{\label{mst}
        Calculated emission spectra from the cavity photonic mode in the cases of emitter-photon coupling induced by (a) none, (b) two, and (c) four phonons.
        The evolution from symmetric to asymmetric lineshape arsing from the phonon-induced coupling is clearly observed.
        (d) Normalized spectra calculated with varying $\gamma_a$.
        The threshold behavior for the asymmetric lineshape is clearly observed.
    }
\end{figure}

The steady state of the cavity QED system is calculated by solving the master equation
\begin{eqnarray}
    \label{master}
    \frac{d}{dt}\rho&=&-\frac{i}{\hbar}[H,\rho]+\sum_{n}{\cal L}(c_n)
\end{eqnarray}
where $\rho$ is the density matrix of the system states and $H$ is Hamiltonian defined in Eq.~(\ref{eqH1}).
Here we consider the system states with one photon and seven phonons allowed at maximum.
The Liouvillian superoperator
\begin{eqnarray}
    \label{decay}
    {\cal L}(c_n) &=& 1/2\left(2c_n\rho c_n^+-\rho c_n^+c_n-c_n^+c_n\rho \right) \ \ \ \ \ \\
    c_{n} &=& \lbrace \sqrt{\gamma_X}\sigma_{G,X}, \sqrt{\gamma_a}a, \sqrt{\gamma_q}q, ,\sqrt{P}\sigma_{X,G} \rbrace \nonumber
\end{eqnarray}
describes Markovian processes corresponding to the decay of emitter, decay of photon, decay of phonon, and pump of emitter with the rate $P$, respectively.
Here we use $\omega_X=0$, $\omega_q=0.1$, $\lambda_{e\text{-}q}=0.2$, $\lambda_{p\text{-}q}=0.1$, $g=0.05$, $\gamma_X=1$, $\gamma_a=0.1$, $\gamma_q=0.1$, $P=0.001$, and a sweeping $\omega_a\in\left(-10,10\right)$ to calculate the emission from the cavity photonic mode.
For brevity, we omit the energy unit meV for the parameters and set $\hbar=1$.
The spectra calculated in the cases of  $k=0,2,4$ are presented in SFig.~\ref{mst}(a)-(c) (two of them already presented in Fig.~1(b)).

We in addition calculate the case with a fixed $\omega_a=0$ and varying $\gamma_a$, and present the normalized spectra in SFig.~\ref{mst}(d).
The threshold behavior for the asymmetric lineshape is clearly observed, that accounts well for the threshold Q-factor presented in Fig.~1(c).
We note that, we use the brief system and parameters around the reference magnitudes \cite{doi:10.1021/acs.nanolett.2c00739,RevModPhys.86.1391} for a qualitative description of the phonon-induced light-matter coupling for $V_B^-$.
The dimensions for the calculated asymmetry (solid line in Fig.~1(c)) are scaled to fit the experimental data.
The actual case could be complex involving multiple phonon modes and many-body effects.
Further simulation is limited by the fact that specific details of $V_B^-$ emission remain unknown yet \cite{Ivady2020,PhysRevB.102.144105,PhysRevLett.128.167401}.
Nonetheless, our brief model already clearly shows how the phonon-induced processes result in the novel optomechanical coupling.

\section{\label{vce} Control Experiments}

\subsection{\label{ppct} Low-Q Cavities}

\begin{figure}
    \includegraphics[width=\linewidth]{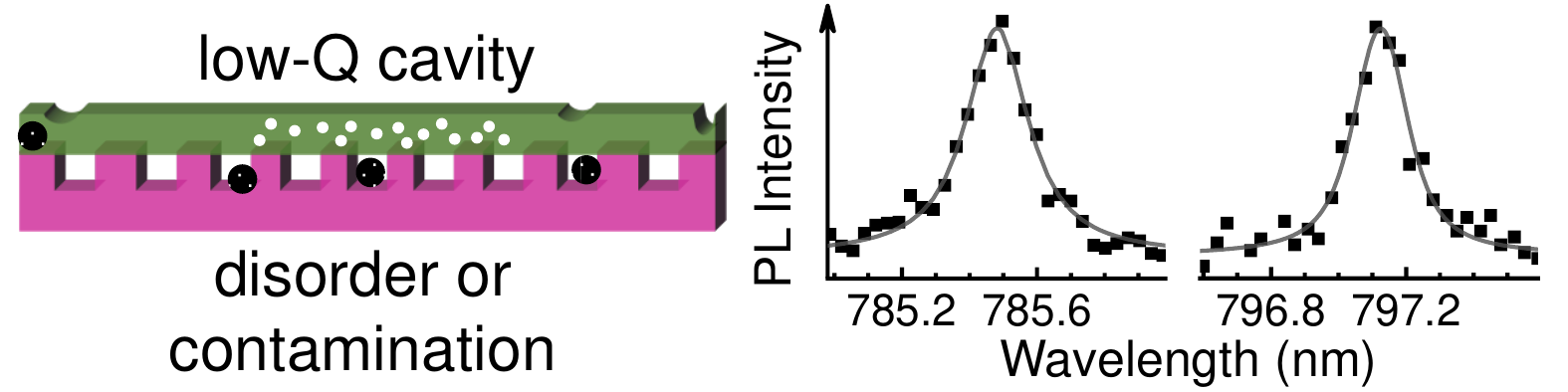}
    \caption{\label{lowq}
        Asymmetry is not observed in irradiated but low-Q ($<10000$) cavities.
    }
\end{figure}

Usually our nanocavity has a high Q-factor $>10^4$.
However, disorder and contamination are sometimes introduced into the cavity during fabrication procedures, which limit the cavity photonic modes to low-Q regime as schematically shown in SFig.~\ref{lowq}.
We do not observe the asymmetric peak for any cavity photonic mode with Q-factor smaller than 10$^4$ such as the two examples on the figure.

\subsection{\label{ram} Identifying Raman Peaks}

\begin{figure}
    \includegraphics[width=\linewidth]{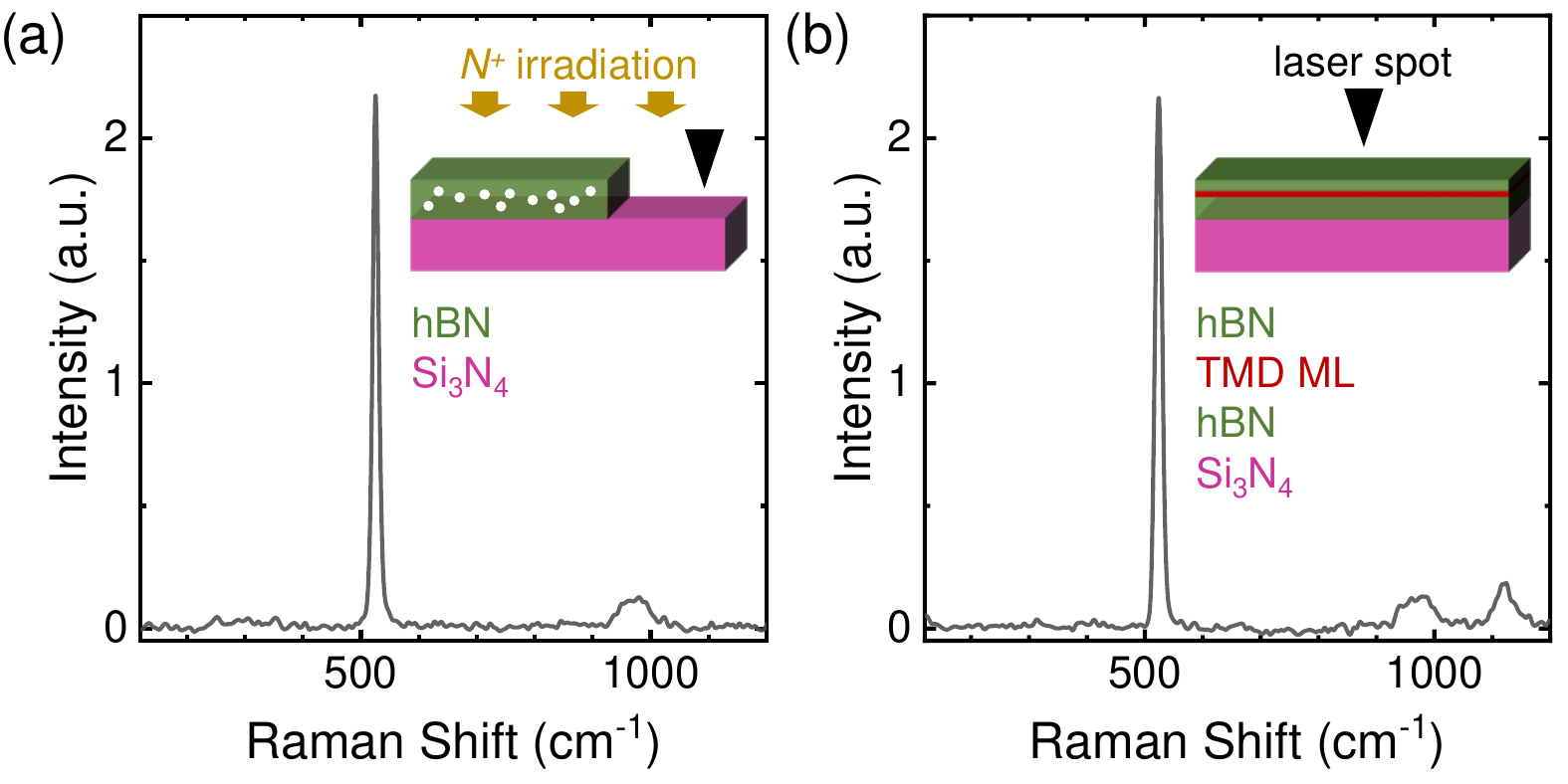}
    \caption{\label{ramc}
        (a) Raman spectrum recorded at Si$_3$N$_4$ substrate after $N^+$ ion irradiation (no hBN).
        (b) Raman spectrum recorded at hBN/MoS$_2$/hBN heterostructur on Si$_3$N$_4$ substrate without irradiation.
        Both are excited by 532nm laser.
    }
\end{figure}

We carry out control experiments to identify the Raman peaks as presented in SFig.~\ref{ramc}.
Raman peaks that have not been observed in these control experiments are thereby identified from $V_B^-$ related phonons.

\subsection{\label{exl} Excluding Optical Reflections}

\begin{figure}
    \includegraphics[width=\linewidth]{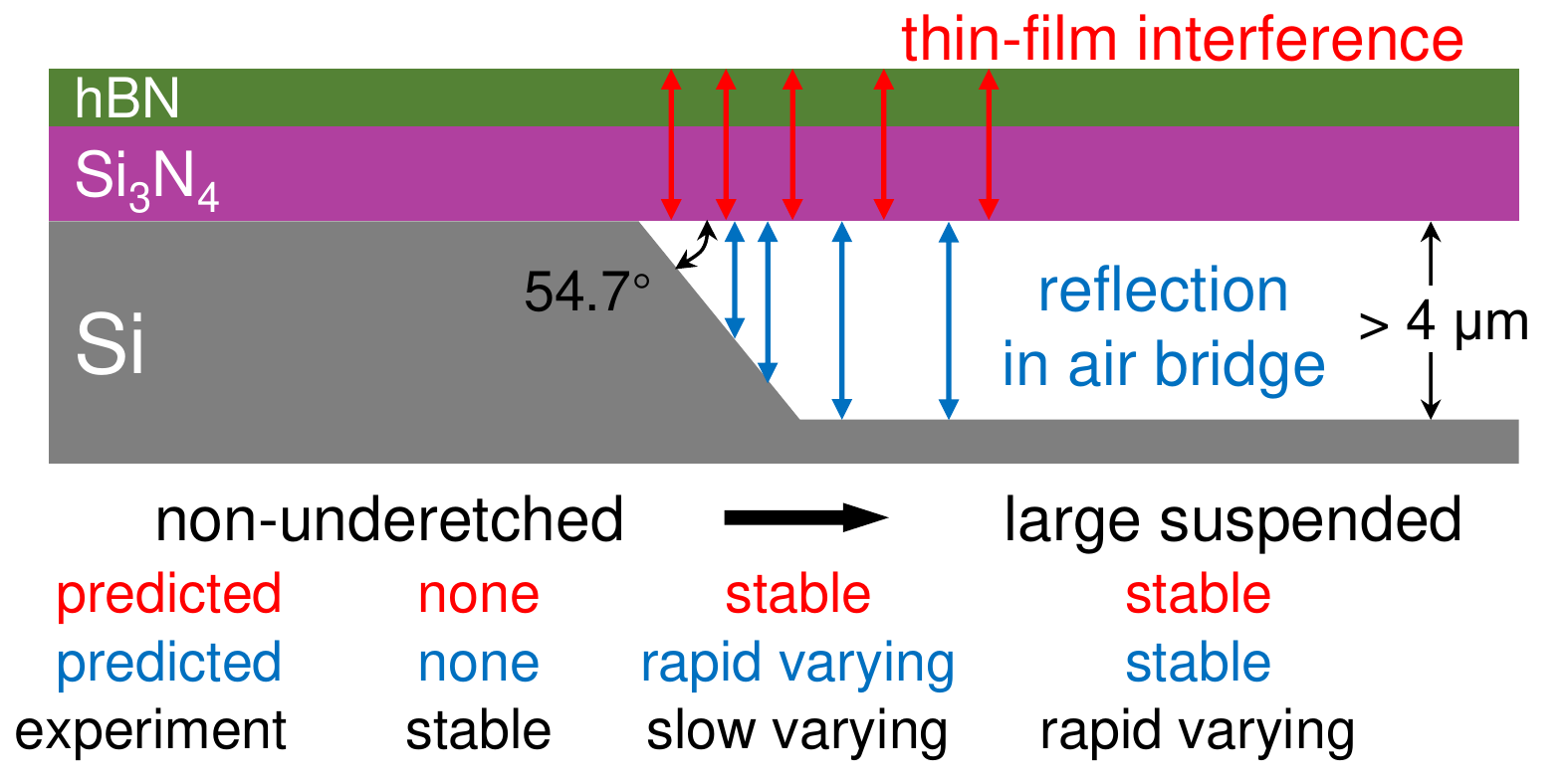}
    \caption{\label{air}
        Excluding air-bridge reflection and thin-film interference.
        As shown, the predicted position dependence of two effects in freely suspended structures can neither match the experimental observations.
    }
\end{figure}

The position and geometry dependence in the PL and Raman spectra of $V_B^-$ centers are well explained by the mechanical modes.
One might wonder if optical reflections e.g., in the air bridge or thin-film shown in SFig.~\ref{air} rather than the mechanical modes result in the position dependence in the spectroscopy.
Hereby, we implement control experiments to exclude these possibilities.

The air bridge in sample-B is etched with 50$\%$ KOH solution, exhibiting an anisotropic etching which means $\langle111\rangle$ surface of Si cannot be etched.
Thus, the etching angle at the boundary of Si is always 54.7$^\circ$ as denoted in SFig.~\ref{air}.
If the optical reflections in the air bridge plays the major role, their effect should vary rapidly at the boundary while be stable away the boundary.
The thin-film interference depends on the thickness of hBN and Si$_3$N$_4$ which have little variation in the whole sample.

Therefore, the position dependence of these two effects can be predicted as shown in SFig.~\ref{air}, but none of them matches our experimental observations.
E.g., as presented in Fig.~3, the position-dependent PL and Raman spectra vary slowly at the boundary while rapidly away from the boundary, opposite to the predictions for reflection effects.
Indeed, due to the high-concentration KOH solution used in the wet underetching, undissolved residuals left at the bottom surface of air bridge as shown in Fig.~2(b).
With such rough surface, the reflection effect in the air bridge is expected to be neglectable.

\subsection{\label{acup} Accuracy in Alighment}

\begin{figure}
    \includegraphics[width=\linewidth]{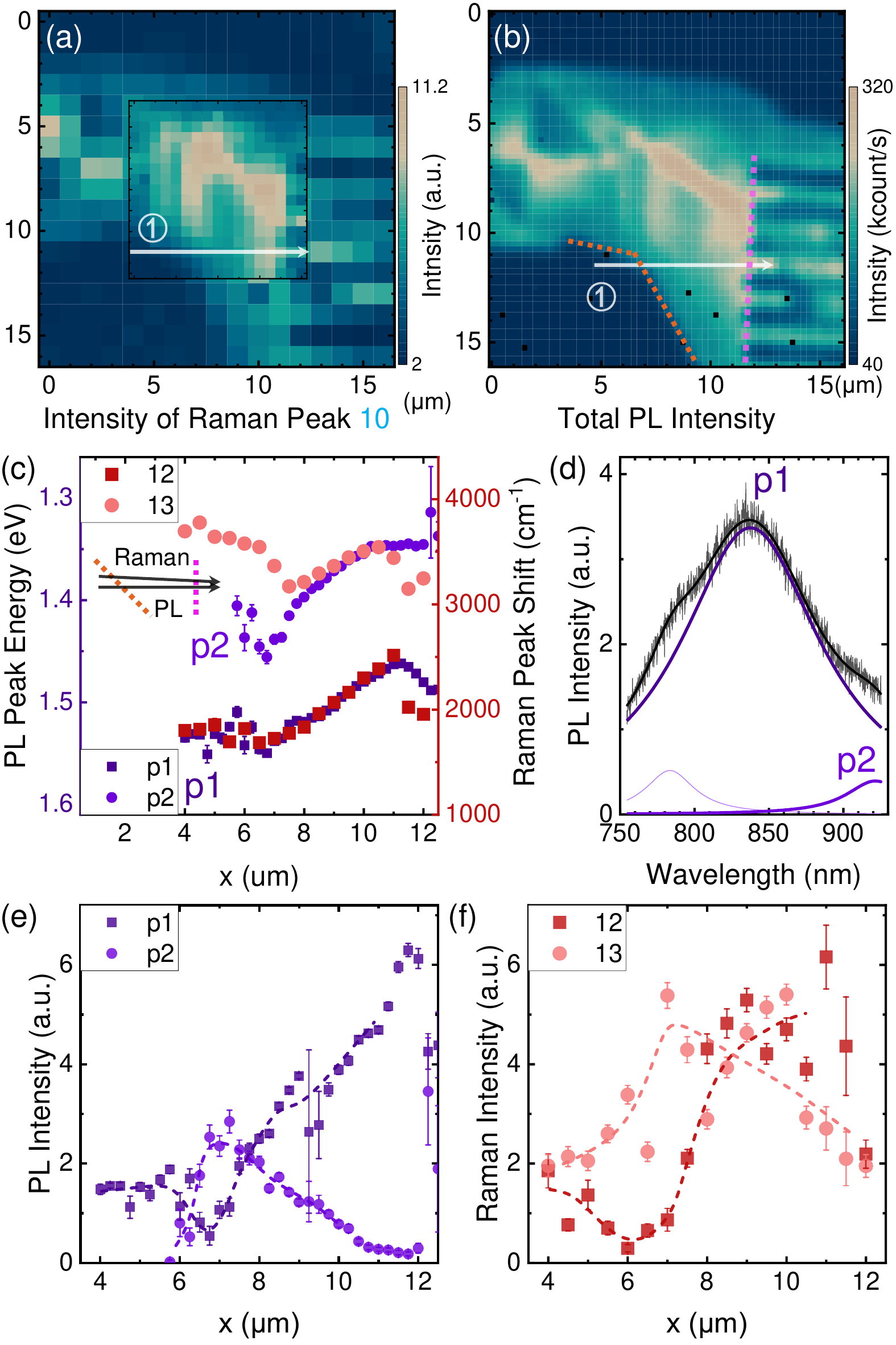}
    \caption{\label{plf2}
        Intensity map of (a) Raman peak 10 of the extended phonon and (b) total $V_B^-$ emission.
        (c) Energies of PL and Raman peaks extracted from trace \textcircled{1}, corresponding to Fig.~3(a)(b).
        Inset depicts potential misalignments between the trace in the two maps.
        (d) Fitting of PL spectrum at $x=12\ \mathrm{\mu m}$.
        Fitting of p1 is accurate whilst that of p2 is limited by the measurement range.
        (e) Intensities of PL peaks and (f) Raman peaks.
        PL peak p1 quantitatively tracks Raman peak 12 whilst PL peak p2 qualitatively tracks Raman peak 13.
    }
\end{figure}

In our measurement, the PL and Raman spectra are recorded separately such as those in SFig.~\ref{plf2}(a)(b).
Thus the accuracy of the comparison between PL and Raman spectra is limited by the misalignments between these two maps.
E.g., some minor mismatch is observed in SFig.~\ref{plf2}(c), which shows the energies of PL and Raman peaks corresponding to Fig.~3(a)(b).
There are mainly two misalignments.
Firstly, the two maps have a small relative twist angle.
Secondly, the maps have different resolutions which result in an unavoidable translational misalignment.
These two misalignments are schematically depicted in the inset in SFig.~\ref{plf2}(c).

Meanwhile, the fitting accuracy of PL peaks at long wavelength is limited by the measurement.
Indeed, spectra in Fig.~3(a)(b) are exactly extracted from the trace \textcircled{1} in SFig.~\ref{plf2}(a)(b).
The PL spectrum at $x=12\ \mathrm{\mu m}$ in SFig.~\ref{plf2}(c) is plotted in (d) along with the multi Lorentz fitting.
As shown, the fitting of PL peak p1 is accurate, but the fitting accuracy of p2 is limited due to the measurement range.
Thus, as shown in SFig.~\ref{plf2}(c), the fitting energy of PL peak p1 quantitatively tracks the shift of corresponding highly localized phonon, whilst that of p2 qualitatively tracks.
The corresponding intensity of PL and Raman peaks are plotted in SFig.~\ref{plf2}(e)(f).
Similarly, the intensity of PL peak p1 agrees quantitatively to that of Raman peak 12, and the intensity of p2 qualitatively agrees to that of Raman peak 13.
This is why we mainly focus on the PL peak p1 when later discussing the superposition of mechanical modes in SFig.~\ref{supp} and Sec.~\ref{cvvm}.
Nonetheless, the two correlations between Raman and PL spectra i.e., the total $V_B^-$ emission intensity follows the intensity of extended phonon peak 10, and the $V_B^-$ emission wavelengths follow the energy of highly localized phonons peak 12 and 13, agree well to the two predicted phonon processes in the $V_B^-$ emission in Fig.~2(a).

\subsection{\label{acros} Anticrossing between $V_B^-$ Phonons}

\begin{figure}
    \includegraphics[width=\linewidth]{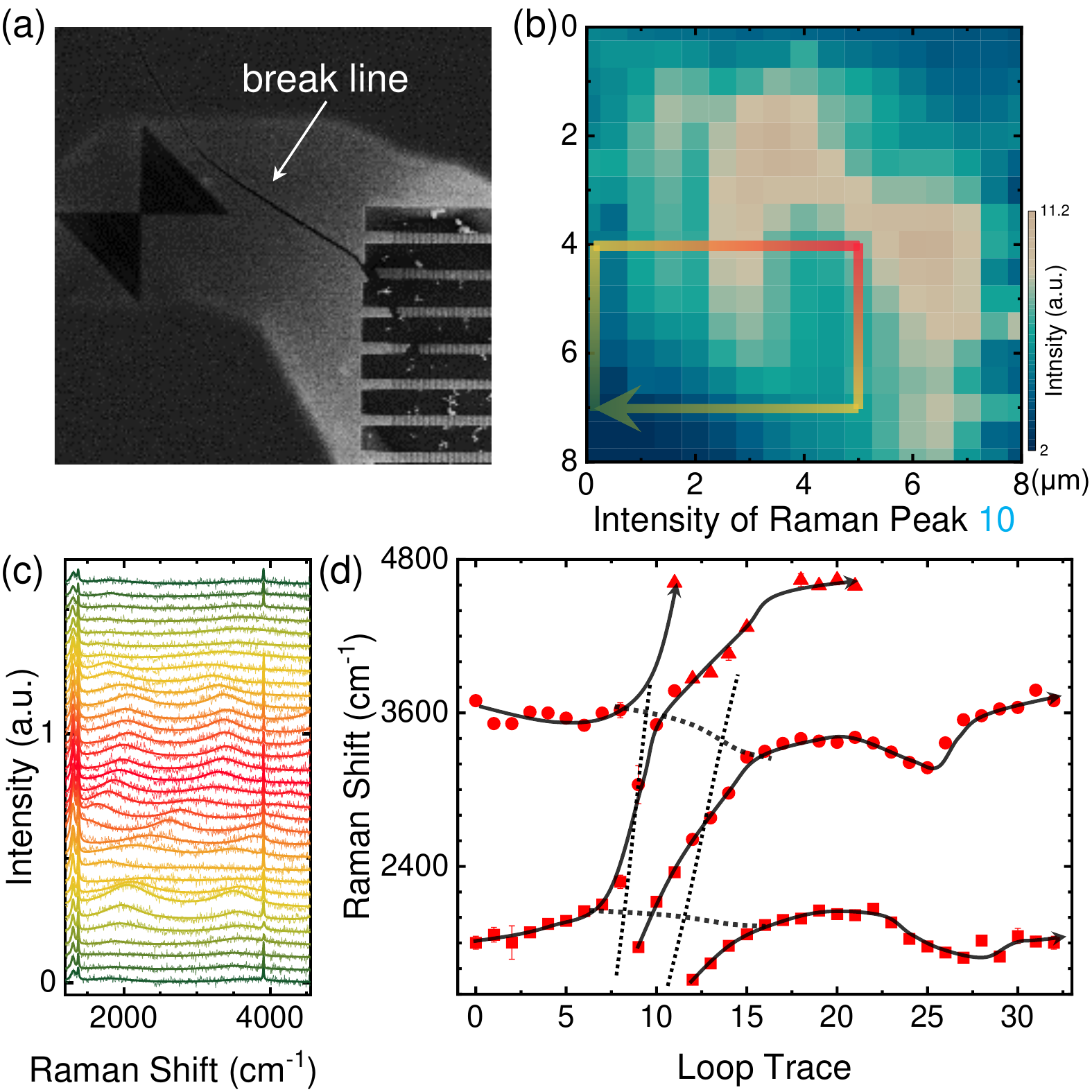}
    \caption{\label{rcp}
        (a) Left top part of the SEM image in Fig.~2(b).
        The large suspended membrane arises from a break line during the sample fabrication.
        (b) Raman map in SFig.~\ref{plf2}(a).
        (c) Raman spectra extracted from the loop trace in (b).
        (d) Energy of highly localized phonons, revealing the anticrossings between Raman peaks.
    }
\end{figure}

The large suspended membrane in the sample indeed arises from a break line during the fabrication as shown in SFig.~\ref{rcp}(a).
In the wet underetching, the KOH solution etches the Si below the large suspended membrane through this break line, which is created during the ICPRIE etching.
The out-of-plane vibration has maximum amplitude around this break line, and thereby, the corresponding total PL intensity and Raman intensity of peak 10 are both strongest, as presented in SFig.~\ref{plf2}(a)(b).

To avoid discontinuous position dependence from the break line, the two traces used in Fig.~3 both avoid the break line.
Here we extract the Raman spectra from the loop trace in SFig.~\ref{rcp}(b) (the break line again avoided) and the results are presented in SFig.~\ref{rcp}(c).
The energy of highly localized phonons are presented in SFig.~\ref{rcp}(d).
Similar to the PL peaks in Fig.~3(c), anticrossings are observed between the Raman spectra as denoted by the dashed lines.
These anticrossings further strengthen the consistence between Raman and PL spectra and support the strong coupling from the Raman point of view.

\subsection{\label{rese} Resonant Excitation}

\begin{figure}
    \includegraphics[width=\linewidth]{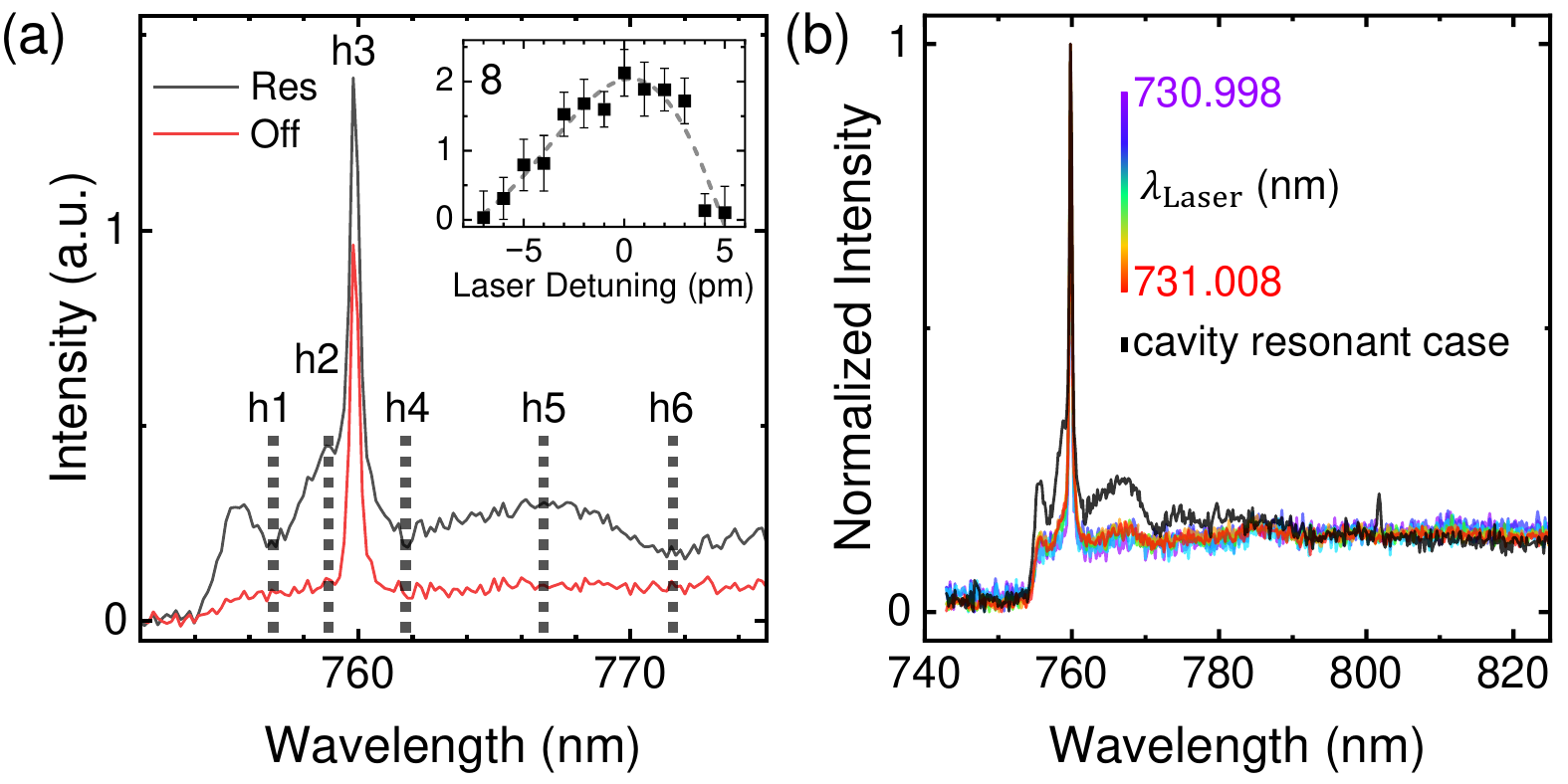}
    \caption{\label{rex}
        (a) Comparison of the spectra at (black) and off (red) resonance.
        Phonon intensities are extracted by the Raman peak heights.
        Inset is the height of Raman peak 8 from $V_B^-$, similar to peak 6 in Fig.~4(b).
        (b) Control experiments. Color lines are normalized spectra recorded at one position in the large suspended membrane (no photonic mode), with the same laser tuned from 730.998 to 731.008 nm.
        Black line is the resonant spectra in (a) for comparison.
    }
\end{figure}

In the resonant excitation of cavity photons presented in Fig.~4, we use a filter to cut the laser signal.
The cutting can be clearly observed in the spectra.
Since the Raman peaks are close to the cut wavelength, their signal might be affected by the filter.
Therefore, we extract the peak height for brevity as shown in SFig.~\ref{rex}(a).
We directly extract the data points as denoted by h1-6.
The height of Raman peak 6 from $V_B^-$ phonon is calculated by $\mathrm{h_{6}}=\mathrm{h}2-\mathrm{h}1$ and peak 7 from Si$_3$N$_4$ by $\mathrm{h_{7}}=\mathrm{h}3-\mathrm{h}2$.
In addition, we extract the height of peak at 766 nm by $\mathrm{h_{8}}=\mathrm{h}5-\left(\mathrm{h}4+\mathrm{h}6\right)/2$.
Similar asymmetric enhancement is again observed for this $V_B^-$ phonon.
We further strengthen the conclusion by the control experiment presented in SFig.~\ref{rex}(b).
We use the same tunable laser and record the spectra at a position in the large suspended membrane (no photonic mode).
As expected, no resonant behavior is observed in the control experiment.
In addition, the enhancement of Si$_3$N$_4$ phonon (Fig.~4(b)) gives a description of the real linewidth (red line) of cavity photonic mode as $3.5 \pm 0.1$ pm, providing a solid evidence of Q-factor $>2\times10^5$ in the ultra-high regime \cite{Song2005}.

\section{\label{fur} Further Specific Explorations}

\subsection{\label{cvvm} Superposition of Cavity Mechanical Modes}

\begin{figure}
    \includegraphics[width=\linewidth]{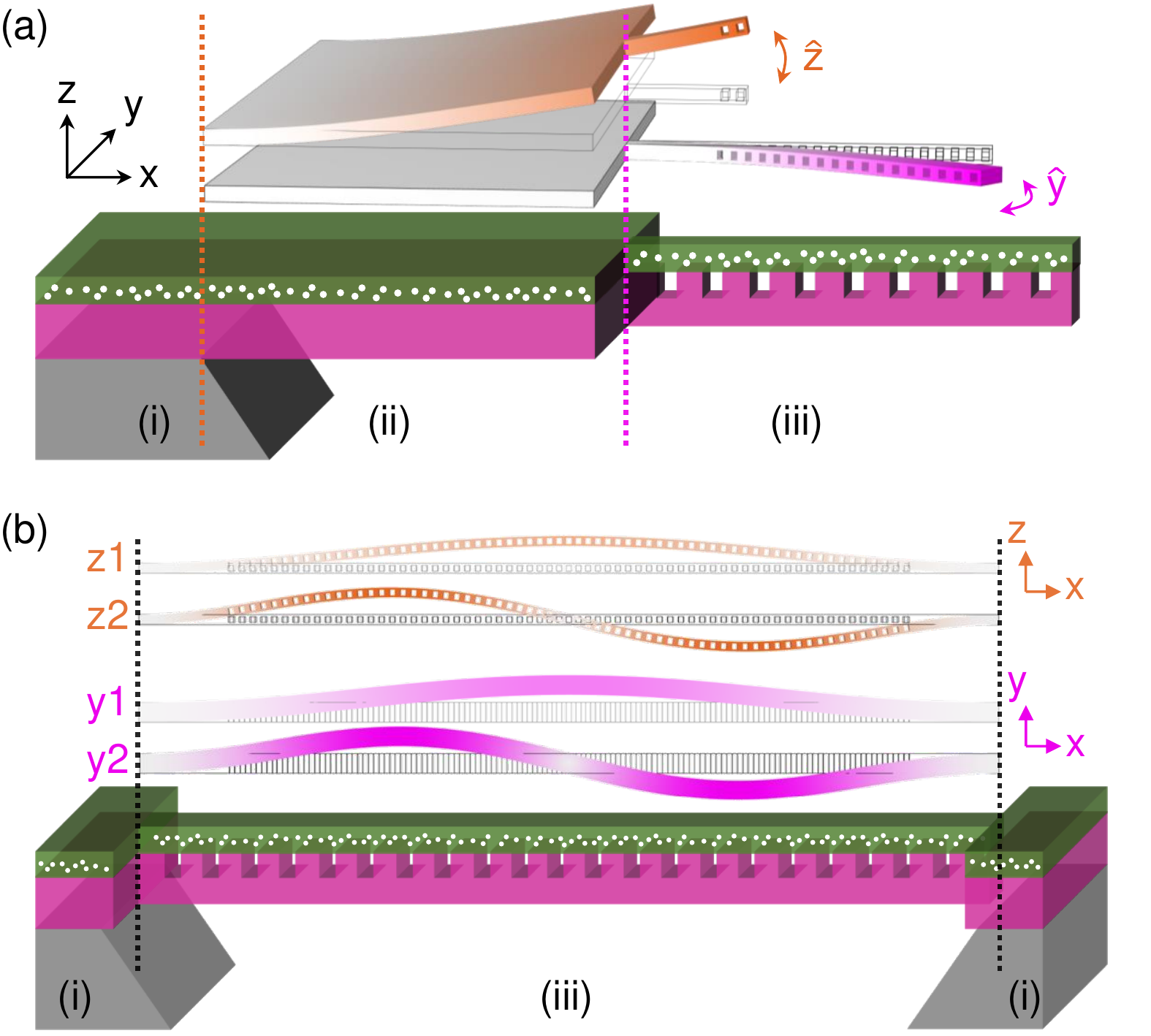}
    \caption{\label{cvs}
        (a) Schematic of the structure involving the large suspended membrane.
        (b) Schematic of the nanobeam directly clamped to the non-underetched region and the corresponding mechanical modes.
    }
\end{figure}

For the case involving the large suspended membrane shown in SFig.~\ref{cvs}(a) (same to Fig.~2(b)), the effects from y-/z- mechanical modes can be separated by the comparison between different regions.
Based on the spatially resolved spectroscopy e.g., data in Fig.~3 and SFig.~\ref{plf2}, we conclude that the out-plane modes in large suspended membrane (ii) enhance the total $V_B^-$ intensity and red shift the PL peak p1, along with the corresponding effects on $V_B^-$ phonons including the enhancement of extended phonon peak 10 and the blue shift of highly localized phonon peak 12.
In contrast, the in-plane modes in the nanobeam (iii) have opposite effects, suppressing the total $V_B^-$ intensity and blue shifts the PL peak p1.
These are the characteristic impacts of in- and out-plane mechanical modes on the $V_B^-$ emission.

For the case where the nanobeam is directly clamped to the non-underetched region shown in SFig.~\ref{cvs}(b), we cannot separate the y-/z- modes spatially.
The state of motion is a superposition of different modes.
We label these modes according to the local harmonic (number of antinodes).
Hereby, the modes are indexed in the figure as y$i$ and z$j$ where $i,j \in (1,2,3...)$ respectively, such as the z1, z2, y1 and y2 modes plotted in SFig.~\ref{cvs}(b).

\begin{figure}
    \includegraphics[width=\linewidth]{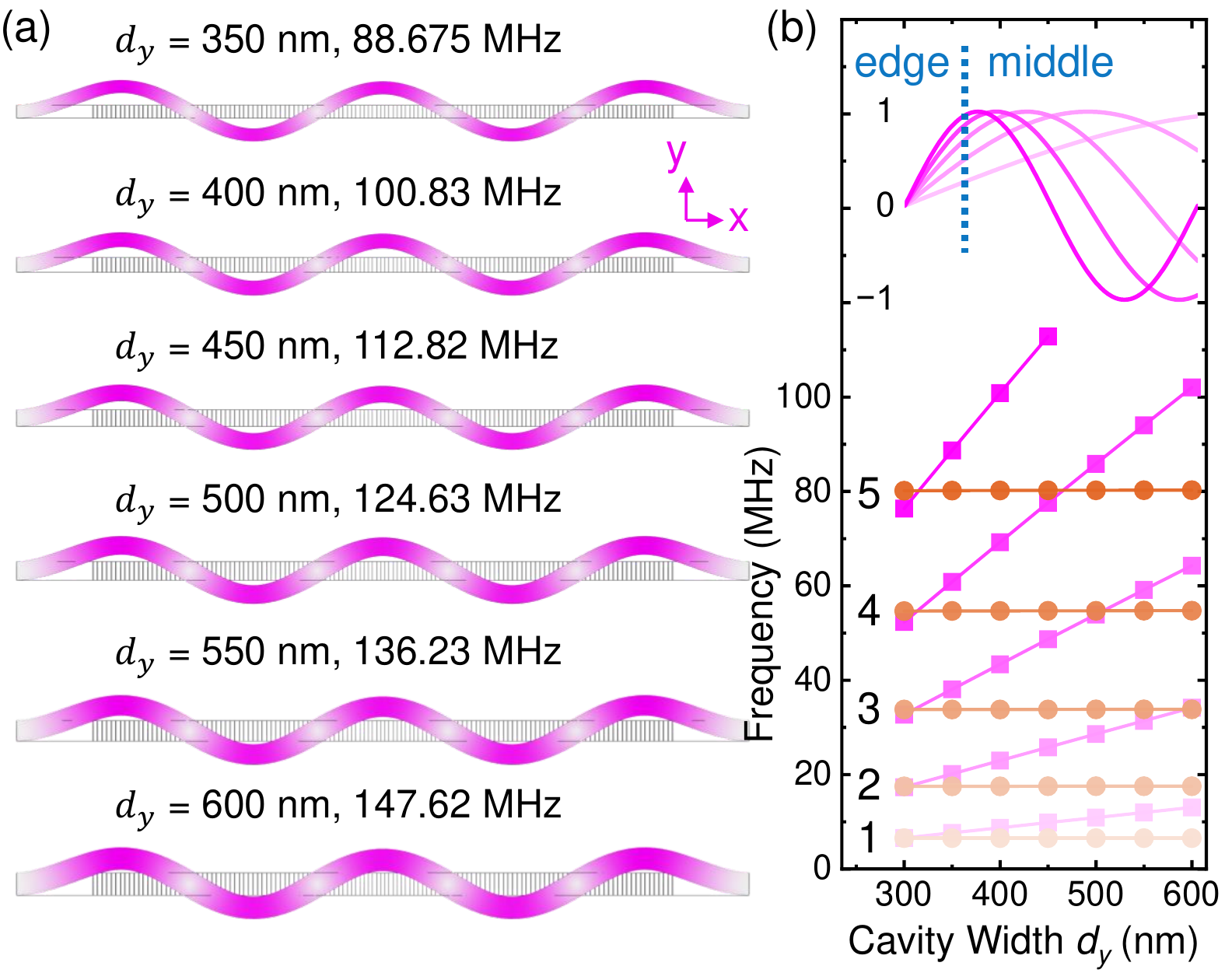}
    \caption{\label{cvv}
        (a) Calculated y5 mode for nanobeam cavities with the varying width $d_y$.
        (b) Calculated vibration frequencies of modes y1-5 and z1-5.
        Inset shows the vibrational profile for modes y1-5.
    }
\end{figure}

We have an array of nanobeam cavities with varying width $d_y$, thus calculate the $d_y$-dependent frequency of y$i$ and z$j$ modes using the finite element method \cite{2204.04304}.
The results are presented in SFig.~\ref{cvv}.
The key result is that the frequency of all y$i$ modes strongly depends on $d_y$, while that of all z$j$ modes rarely changes with $d_y$.
We note that the optical and mechanical properties of hBN are not well known yet and strongly depend on the specific flake e.g. the local defect concentration \cite{Thomas_2015}.
For brevity, we tentatively use the parameters of Si$_3$N$_4$ for hBN, to calculate numerical results \cite{2204.04304}.
This might introduce differences between the theory and experiment.
However, we emphasize that the varying y$i$ frequencies are due to the bending rigidity in y direction changes with the spatial extent $d_y$ \cite{doi:10.1021/acs.nanolett.2c00613}.
In contrast, the bending rigidity in z direction rarely changes, due to the spatial extent in z direction (thickness of hBN and Si$_3$N$_4$) is same for all nanobeams.
The geometric dependence of bending rigidity is a fundamental property in mechanics and does not rely on the parameters of specific materials.

\begin{figure}
    \includegraphics[width=\linewidth]{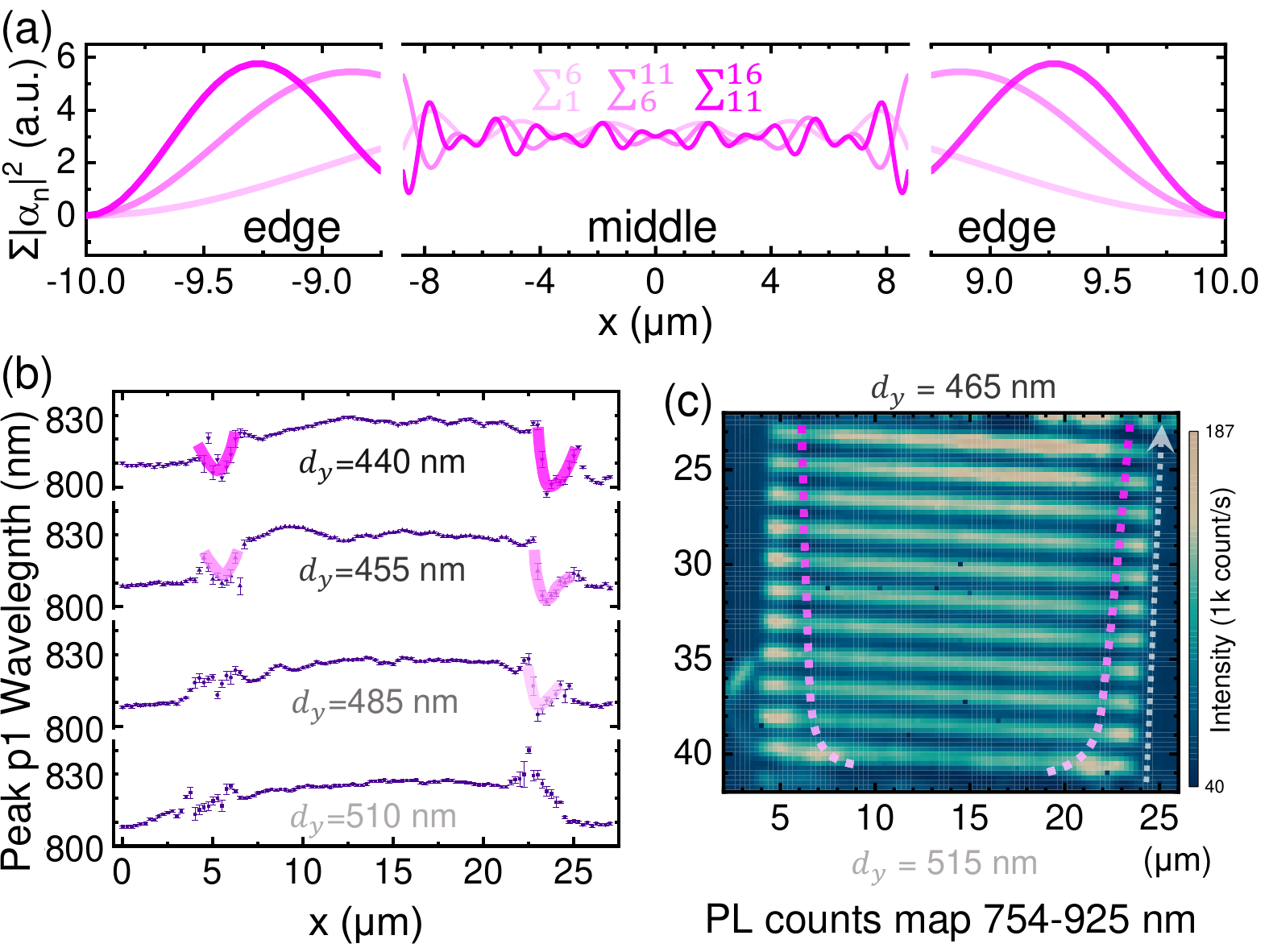}
    \caption{\label{supp}
        (a) The sum amplitude of multiple y$i$ modes.
        $\sum_{1}^{6}$, $\sum_{6}^{11}$ and $\sum_{11}^{16}$ are the simulation of a broad, medium and narrow cavity, respectively.
        (b) PL shift of p1 and (c) total PL intensity in nanobeam cavities, corresponding to the bottom part of the SEM image in Fig.~2(b).
        Magenta lines in (b)(c) denote the blue shift of p1 and the suppression of total PL intensity at nanobeam edges.
        These are both y-effects and increase as $d_y$ decreases, agreeing perfectly to the theoretical predictions in (a).
    }
\end{figure}

Next, we use the simple trigonometric profiles
\begin{eqnarray}
    \label{triv}
    \alpha_i=\sin\left(i\pi\left(x/L+1/2 \right)\right)
\end{eqnarray}
to describe the eigenmodes and predict the superposition of multiple y$i$ modes.
Here, $L$ is the nanobeam length and $x\in\left(-L/2,L/2\right)$ is the position relative to the center point \cite{Khaniki2022}.
We simulate the interaction effect of the superposition state to $V_B^-$ by summing the amplitude of multiple modes with random phase factors.
Strictly, the summed coupling effect $a\left(x\right)$ should be
\begin{eqnarray}
    \label{sums}
    a\left(x\right)=\sum_{f} n\left(f\right) \vert g\left(f\right) \alpha_f \left(x\right) \vert^2
\end{eqnarray}
, where $\alpha_f$ is the vibration profile of mode y$i$ which has the frequency $f$, $n\left(f\right)$ is the Bose factor, and $g\left(f\right)$ reflects the coupling strength to $V_B^-$.
For a given $f$, the corresponding mode order $i$ changes with $d_y$.
$g\left(f\right)$ is an intrinsic property of $V_B^-$ and should not change with $d_y$.
Theoretically we could predict the $d_y$-dependent superposition based on Eq.~(\ref{sums}).

However, at this point $g\left(f\right)$ is difficult to get by either calculation or experiment.
For a simple but qualitative example, we calculate the superposition according to
\begin{eqnarray}
    \label{suml}
    a\left(x\right)=\sum_{i\in G} \vert \alpha_i \left(x\right) \vert^2
\end{eqnarray}
where $i\in G$ means the y$i$ modes have the frequencies within the range $G$ which couples to $V_B^-$.
$G$ is an intrinsic property of $V_B^-$, and the corresponding harmonics $i$ within $G$ changes with $d_y$.
We assume that for a broad cavity with large $d_y$, $i\in G=1-6$ thus the superposition (coupling) $a\left(x\right)=\sum_1^6$.
Similarly, for a medium cavity $i\in G=6-11$ and for a narrow cavity  $i\in G=11-16$.
The three superpositions are presented in Fig.~\ref{supp}(a).
As shown, $a\left(x\right)$ at nanobeam edge clearly increase as $d_y$ decreases.
This is due to the nanobeam edge $x=\pm L/2$ is the node of all modes, thus the sum of multiple y$i$ modes retains the $d_y$ dependence of each single mode as presented in the inset of SFig.~\ref{cvs}(b).
In contrast, in the middle region, the nodes of different modes are interlaced, and thereby, the sum exhibits only minor variations.

The highly simplified approach Eq.~(\ref{suml}) is not strictly equal to the correct approach Eq.~(\ref{sums}).
However, we can expect the result obtained by Eq.~(\ref{sums}) would be qualitatively similar to the simple approach Eq.~(\ref{suml}), because the two fundamental properties -- the nanobeam edge is the node of all modes and the nodes of multiple modes are interlaced in the middle -- are always correct and not affected in Eq.~(\ref{suml}).
Indeed, the theoretical predictions in SFig.~\ref{supp}(a) agree perfectly to the experimental observations as presented in SFig.~\ref{supp}(b)(c).
As discussed in the context of SFig.~\ref{plf2}, y-vibrations blue shift the PL peak p1 and suppress the PL intensity.
The "concave" blue shift of PL peak p1 in SFig.~\ref{supp}(b), corresponding to the "convex" y-amplitude predicted in SFig.~\ref{supp}(a), becomes significant as $d_y$ decreases.
Meanwhile, from edge to middle the PL intensity is firstly enhanced (z-effect) and then suppressed (y-effect), resulting in the bright bar at nanobeam edges shown in SFig.~\ref{supp}(c).
As $d_y$ decreases, the bright bar becomes shorter, indicating stronger suppression from y-vibrations.
In contrast to these clear effects at nanobeam edges, the $V_B^-$ emission in the middle region are generally stable but with minor variations.
All these results are consistent to theoretical predictions in SFig.~\ref{supp}(a).
This perfect agreement of $d_y$ dependence further supports our conclusions in the emitter-optomechanical interaction.

Besides the vibration of whole nanobeam presented in SFig.~\ref{supp}, our cavity also supports the confined mechanical modes ($\sim$ GHz) at the cavity center \cite{10.1038/nature08524,2204.04304}.
We previously found that the MoS$_2$ exciton couples to both the beam and confined mechanical modes \cite{2204.04304}.
In contrast, in most cavities here no particular change of $V_B^-$ emission wavelength or intensity is observed at center as shown in SFig.~\ref{supp}(b)(c).
This result might be due to that $V_B^-$ prefers to couple to MHz rather than GHz modes, or the high-frequency modes are degraded by the air (SFig.~\ref{wow}).

\subsection{\label{pshft} Other $V_B^-$-Related Phonons}

\begin{figure}
    \includegraphics[width=\linewidth]{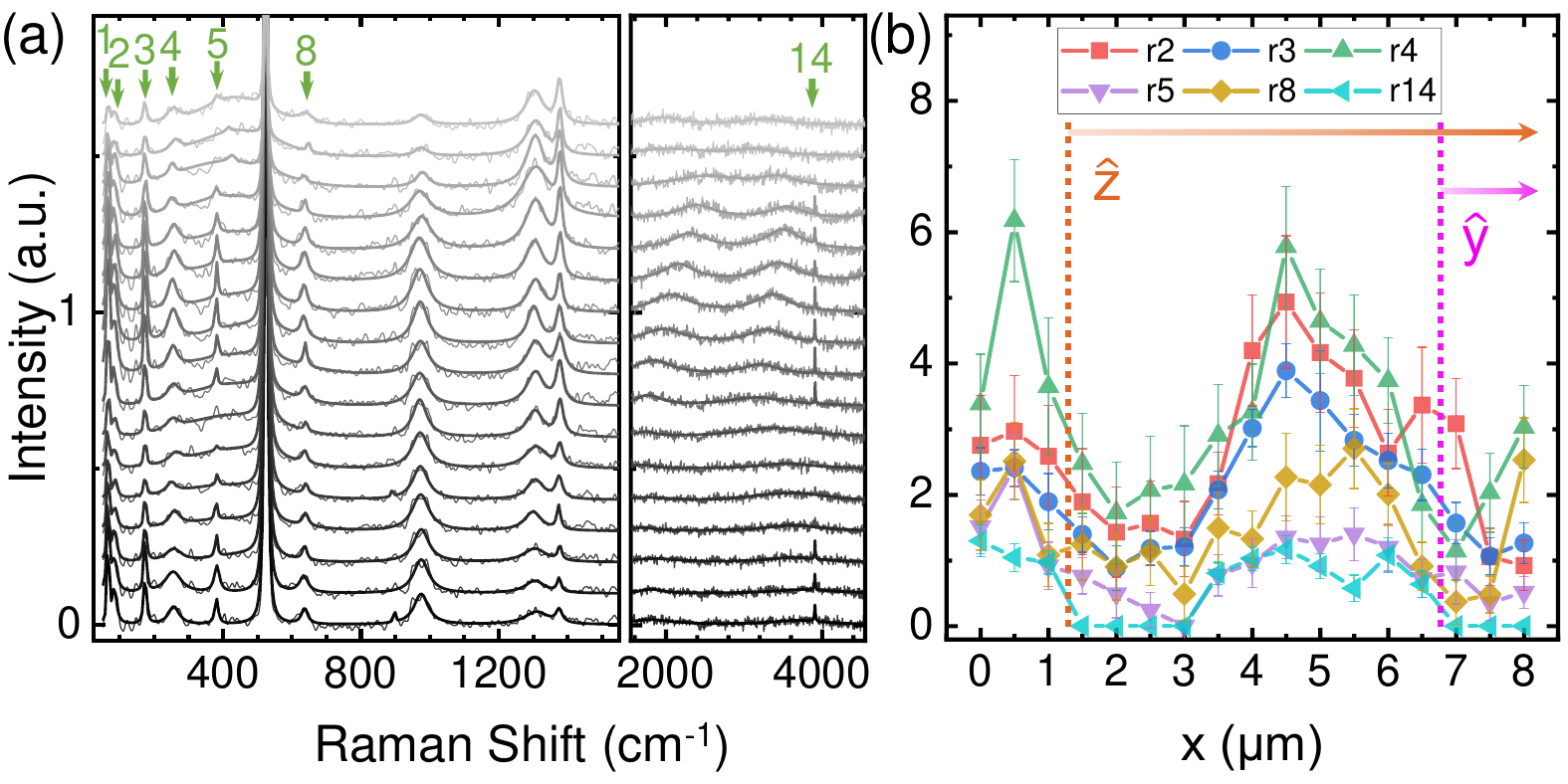}
    \caption{\label{grp}
        Raman peaks 1-5, 8 and 14 exhibit few direct relation to the $V_B^-$ emission.
        (a) Position-dependent Raman spectra, where the right panel is already presented in Fig.~3.
        (b) Their intensity variation.
    }
\end{figure}

We present the whole-range Raman spectra of Fig.~3(a) in SFig.~\ref{grp}(a) here.
The intensity of green labelled peaks 1-5, 8 and 14 are presented in SFig.~\ref{grp}(b), along with the boundaries between the regions (i)-(iii) (orange and magenta dashed lines) denoted.
As shown, these peaks exhibit the same position dependence: a suppression at both two boundaries.
Meanwhile, they have the energy 67, 86, 174, 256, 382, 638 and 3909 $\mathrm{cm^{-1}}$ respectively, and no distinguishable shift is observed.
Comparing to the $V_B^-$ emission (Fig.~3(b)), we find no direct relation of these phonons to either the intensity or the wavelengths of $V_B^-$ emission.
G. Grosso et al. \cite{doi:10.1021/acsphotonics.9b01789} recently reported the investigation on the phonon processes for several nitrogen vacancies in hBN.
They found that for nirogen vacancies, phonons around $150-200\ \mathrm{meV}$ dominate the excitation of emitters and determines the total PL intensity.
We observe the same phenomenon for $V_B^-$ that, the extended phonon peak 10 at $162\ \mathrm{meV}$ follows the total emission intensity of $V_B^-$, as presented in SFig.~\ref{plf2}(a)(b).
Meanwhile, G. Grosso et al. also found the low-energy acoustic and out-plane optical phonons play few roles in emission of nitrogen vacancies \cite{doi:10.1021/acsphotonics.9b01789}.
Therefore, we suggest that Raman peaks 1-5, 8 and 14 are the acoustic or out-plane optical phonons of $V_B^-$.

We note that some specific details could be interesting future topics.
Considering the zero-phonon emission of $V_B^-$ at $\sim773$ nm \cite{doi:10.1021/acs.nanolett.2c00739} and the replica peak p1 at $\sim805$ nm in the non-underetched region, we can expect that peak p1 arises from the zero-phonon state by firstly absorbing an extended phonon peak 10 ($1306\ \mathrm{cm^{-1}}$) and then emit a highly localized phonon peak 12 ($1800\ \mathrm{cm^{-1}}$).
In addition, although no direct relation with the $V_B^-$ emission is observed, the green labelled phonons also exhibit interaction effect with the vibrations i.e., the intensity variation as presented in SFig.~\ref{grp}.
These results indicate specific features in the emission dynamics of $V_B^-$ are not well studied yet.
We suggest that these features might be related to the hyperbolic dispersion of hBN \cite{PhysRevLett.109.104301} since it provides a strong confinement of phonons which could improve their coupling to electronic transitions.
Such hyperbolic features can be controlled by nanostructures \cite{Caldwell2014,Pons-Valencia2019} or even a suspending of hBN flakes \cite{doi:10.1021/acs.nanolett.8b04242}, similar to the cavity-induced control of $V_B^-$ centers we observed.

\clearpage